\documentclass[a4paper,11pt]{article}
\usepackage{pos}

\usepackage{latexsym}

\newcommand{\be}{\begin{eqnarray}}
\newcommand{\ee}{\end{eqnarray}}

\title{Testing General Relativity\\with Black Hole X-Ray Data and ${\tt ABHModels}$}

\ShortTitle{Testing General Relativity with Black Holes}

\author*[a]{Cosimo~Bambi}
\author[b,a,c]{Askar~B.~Abdikamalov}
\author[a]{Honghui~Liu}
\author[a,d]{Shafqat~Riaz}
\author[a]{Swarnim~Shashank}
\author[e]{Menglei~Zhou}

\affiliation[a]{Center for Field Theory and Particle Physics and Department of Physics, Fudan University,\\ 
200438 Shanghai, China}

\affiliation[b]{School of Mathematics and Natural Sciences, New Uzbekistan University,\\
Tashkent 100007, Uzbekistan}

\affiliation[c]{Ulugh Beg Astronomical Institute,\\
Tashkent 100052, Uzbekistan}

\affiliation[d]{Theoretical Astrophysics, Eberhard-Karls Universit\"at T\"ubingen,\\
D-72076 T\"ubingen, Germany}

\affiliation[e]{Institut f\"ur Astronomie und Astrophysik, Eberhard-Karls Universit\"at T\"ubingen,\\
D-72076 T\"ubingen, Germany}

\emailAdd{bambi@fudan.edu.cn}

\abstract{The past 10~years have seen tremendous progress in our capability of testing General Relativity in the strong field regime with black hole observations. 10~years ago, the theory of General Relativity was almost completely unexplored in the strong field regime. Today, we have gravitational wave data of the coalescence of stellar-mass black holes, radio images of the supermassive black holes SgrA$^*$ and M87$^*$, and high-quality X-ray data of stellar-mass black holes in X-ray binaries and supermassive black holes in active galactic nuclei. In this manuscript, we will review current efforts to test General Relativity with black hole X-ray data and we will provide a detailed description of the public codes available on ${\tt ABHModels}$.}

\FullConference{
Multifrequency Behaviour of High Energy Cosmic Sources XIV (MULTIF2023)\\
12-17 June 2023\\
Palermo, Italy\\}

\begin{document}
\maketitle

\section{Introduction}

General Relativity is one of the pillars of modern physics. The theory was proposed by Einstein at the end of 1915~\cite{Einstein:1916vd} and has successfully passed a large number of observational tests without requiring any modification from its original version. General Relativity has been extensively tested in the weak field regime with experiments in the Solar System and radio observations of binary pulsars~\cite{Will:2014kxa}. In the past 20~years, there have been significant efforts to test the theory on large scales with cosmological observations~\cite{Ferreira:2019xrr}. More recently, we have started testing General Relativity even in the strong field regime with black holes. Testing General Relativity with black hole data is a new and promising line of research, which is possible today -- and was not possible only 10~years ago -- thanks to a new generation of observational facilities.

Black holes are the sources of the strongest gravitational fields that can be found today in the Universe and are thus the best laboratories for testing General Relativity in the strong field regime~\cite{Bambi:2017khi}. In General Relativity, black holes are relatively simple objects, as they are completely characterized by a small number of parameters. This is the celebrated result of the no-hair theorem, which is actually a family of theorems with different versions and a number of extensions~\cite{Carter:1971zc,Robinson:1975bv,Chrusciel:2012jk}. The spacetime metric around astrophysical black holes is though to be described well by the Kerr solution~\cite{Kerr:1963ud} and be completely characterized by two parameters, which are associated to the mass $M$ and the spin angular momentum $J$ of the object\footnote{Deviations from the Kerr solution from ``standard'' physics (non-vanishing electric charge of the object, presence of an accretion disk or nearby stars, etc.) can be quantified, but they turn out to be completely negligible in normal situations; see, for instance, Refs.~\cite{Bambi:2017khi,Bambi:2008hp,Bambi:2014koa} and references therein.}. On the other hand, macroscopic deviations from the Kerr spacetime are possible if General Relativity is not the correct theory of gravity~\cite{Yunes:2009hc,Kleihaus:2011tg}, in models with macroscopic quantum gravity effects at the black hole event horizon~\cite{Dvali:2011aa,Giddings:2014ova,Mottola:2023jxl}, and in the presence of exotic matter fields~\cite{Herdeiro:2014goa,Herdeiro:2016tmi}.

Today there are at least three leading methods for testing General Relativity in the strong field regime with black holes~\cite{Bambi:2015kza,Yagi:2016jml}: the study of the gravitational wave signal emitted by the coalescence of two stellar-mass black holes with gravitational wave laser interferometer data~\cite{LIGOScientific:2016lio,Yunes:2016jcc,LIGOScientific:2019fpa}, the study of the images of the supermassive black holes SgrA$^*$ and M87$^*$ with mm VLBI data~\cite{Bambi:2019tjh,EventHorizonTelescope:2020qrl,Vagnozzi:2022moj}, and the analysis of the spectra of accreting black holes with X-ray data~\cite{Cao:2017kdq,Tripathi:2018lhx,Tripathi:2020yts}. Gravitational wave tests and black hole imaging tests are quite popular nowadays and there is a rich literature on the two subjects. X-ray tests are definitively less popular than the other two methods. However, as we will show in the next sections, X-ray tests can provide stringent and competitive constraints on possible deviations from the Kerr solution in the strong gravity region of black holes. There are already two recent reviews on tests of General Relativity with black hole X-ray data~\cite{Bambi:2022dtw,Bambi:2023mca}, and the interest reader can find many details about this line of research there. In particular, in Ref.~\cite{Bambi:2022dtw} there is a detailed discussion about the systematic effects and the accuracy of these tests. In the present manuscript, we will briefly review the results and we will instead describe the details of the codes used for these tests. These codes are public on GitHub\footnote{\href{https://github.com/ABHModels}{https://github.com/ABHModels}}, but a tutorial has never been published so far.

The manuscript is organized as follows. In Section~\ref{s-dcm}, we will briefly review the astrophysical systems for our tests with X-ray data. In Section~\ref{s-ss}, we will outline the main steps to calculate thermal and reflection spectra of accretion disks. In Section~\ref{s-models}, we will describe the public models available on ${\tt ABHModels}$. In Section~\ref{s-results}, we will briefly review the main results of our tests of General Relativity with black hole X-ray data. Summary and conclusions are reported in Section~\ref{s-fr}. Throughout the manuscript, we will employ natural units in which $c = G_{\rm N} = \hbar = k_{\rm B} = 1$.


\section{Disk-Corona Model}\label{s-dcm}

Our tests of General Relativity in the strong field regime with black hole X-ray data require very special astrophysical systems. From stellar evolution simulations, we expect around 10$^8$~stellar-mass black holes in a galaxy like the Milky Way~\cite{Timmes:1995kp,Olejak:2019pln}. Despite such a huge number of black holes around us, today we know only around 70~stellar-mass black holes from X-ray observations (they are almost all in the Milky Way and only a few of them in nearby galaxies). Since our tests of General Relativity require high-quality spectra with specific properties~\cite{Bambi:2022dtw}, in the end we have only a few observations of a few sources suitable for our tests!

\begin{figure}[h!]
\begin{center}
\includegraphics[width=10.5cm]{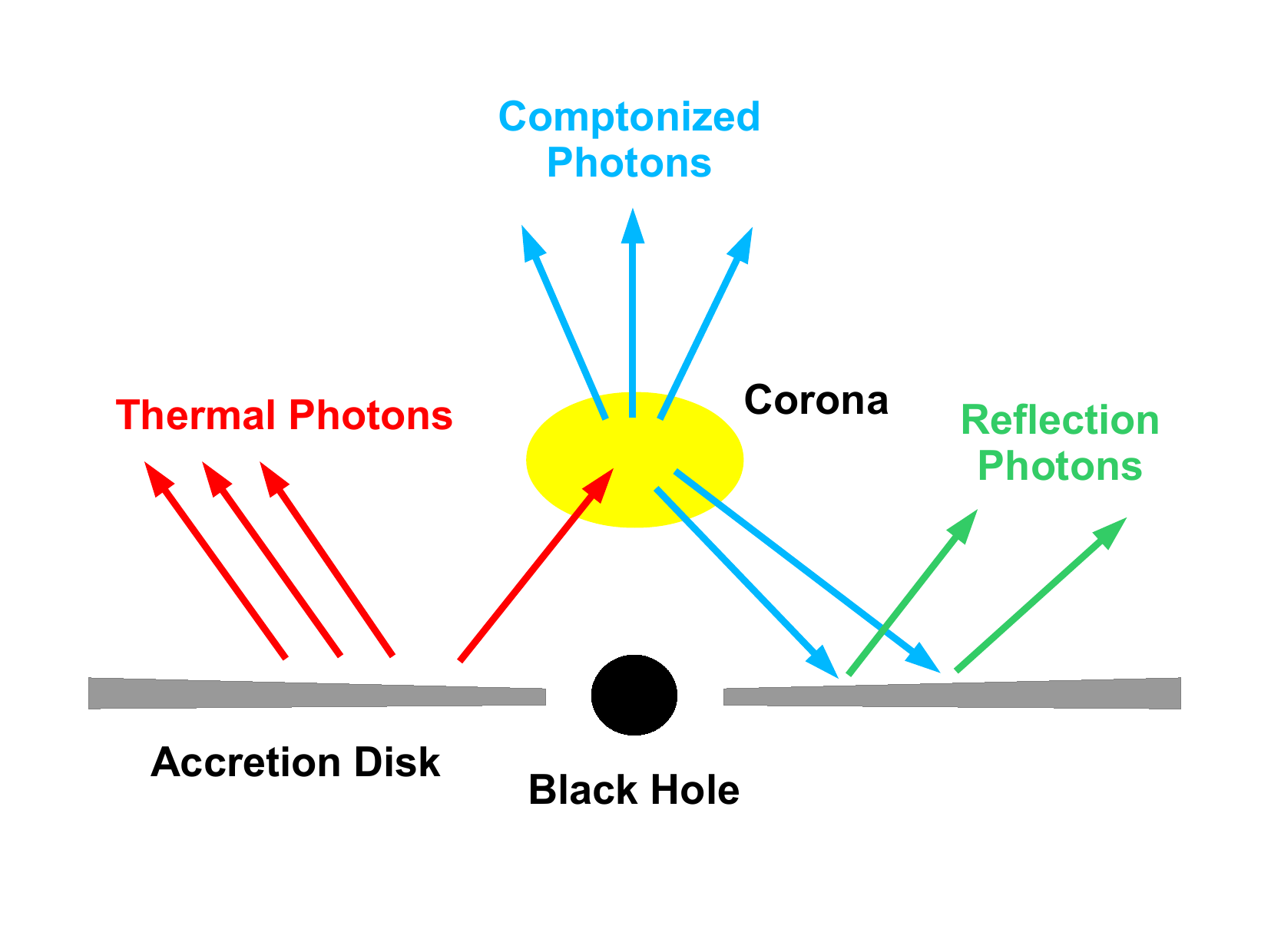}
\end{center}
\vspace{-1.4cm}
\caption{Disk-corona model: the black hole is accreting from a cold, geometrically thin, and optically thick disk and there is a hot corona enshrouding the black hole and the inner part of the disk. Thermal photons from the disk inverse Compton scatter off free electrons in the corona. Some Comptonized photons illuminate the disk and produce the reflection component. Figure from Ref.~\cite{Bambi:2021chr} under the terms of the Creative Commons Attribution 4.0 International License. \label{f-corona}}
\vspace{0.5cm}
\begin{center}
\includegraphics[width=14.0cm]{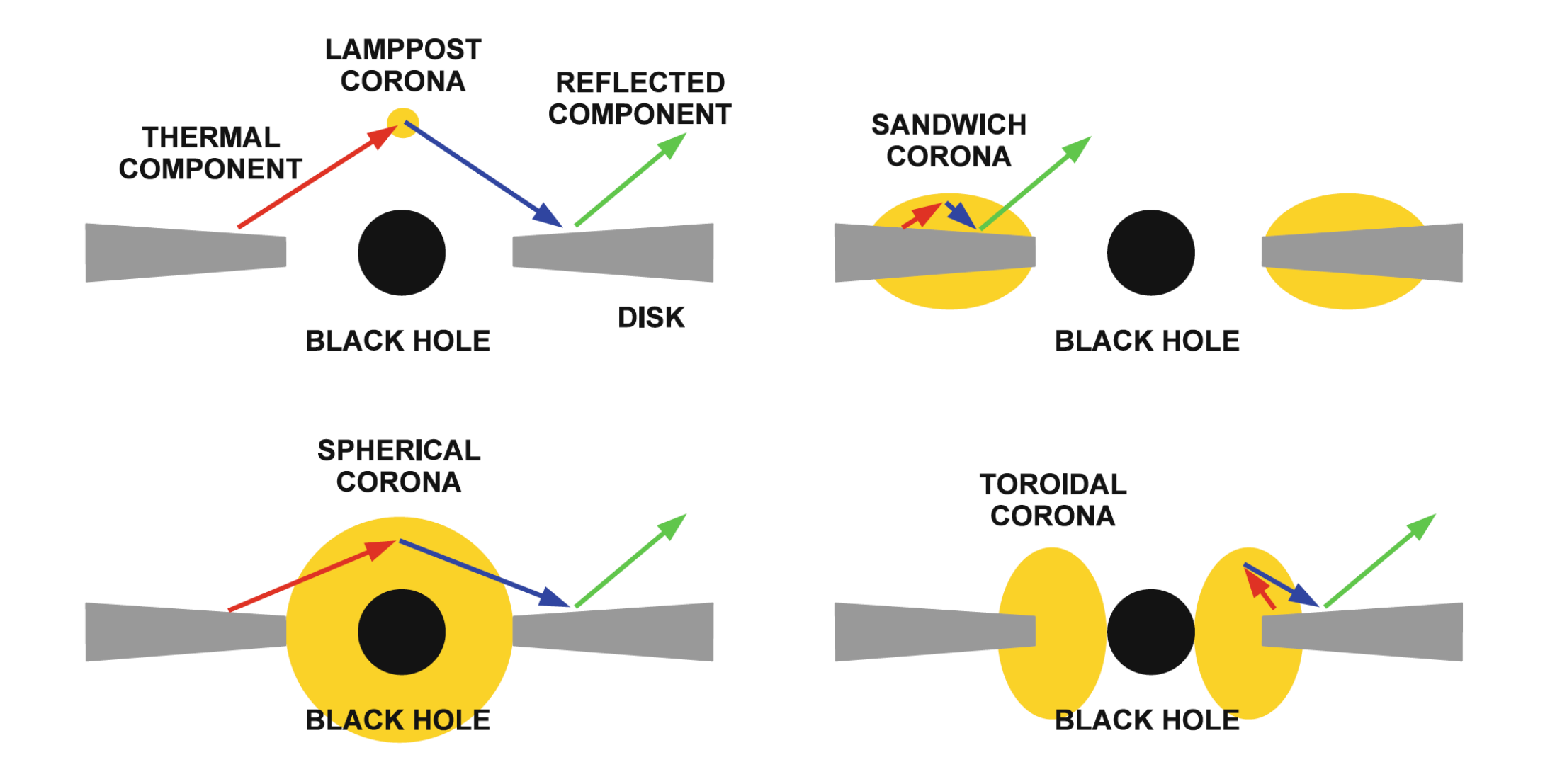}
\end{center}
\vspace{-0.4cm}
\caption{Coronal geometries: lamppost corona (top left panel), sandwich corona (top
right panel), spherical corona (bottom left panel), and toroidal corona (bottom right panel). Figure from Ref.~\cite{Bambi:2017khi}. \label{f-corona2}}
\end{figure}

The astrophysical system of our tests is shown in Fig.~\ref{f-corona} and is normally referred to as the disk-corona model~\cite{Bambi:2020jpe}. The black hole can either be a stellar-mass black hole in an X-ray binary or a supermassive black hole in an active galactic nucleus. The crucial point is that the black hole is accreting from a cold, geometrically thin, and optically thick disk. In such a condition, every point on the surface of the disk emits a blackbody-like spectrum and the whole disk has a multi-temperature blackbody-like spectrum, because the temperature of the material in the disk increases as we approach the black hole. The thermal spectrum of the disk is normally peaked in the soft X-ray band ($\sim 1$~keV) for stellar-mass black holes and in the UV band (1-100~eV) for supermassive black holes~\cite{Bambi:2017khi}. The {\it corona} is some hot plasma ($\sim 100$~keV) near the black hole and the inner part of the accretion disk: the corona may be the base of the jet (lamppost corona), the hot atmosphere above the accretion disk (sandwich corona), or the material in the plunging region between the inner edge of the accretion disk and the black hole (spherical and toroidal coronae), see Fig.~\ref{f-corona2}.

Since the accretion disk is cold and the corona is hot, thermal photons from the accretion disk can inverse Compton scatter off free electrons in the corona. The Comptonized photons have a spectrum in the X-ray band that can be normally approximated well by a power law with an high-energy cutoff, where the high-energy cutoff at the corona should be close to the coronal temperature (while the high-energy cutoff in the X-ray spectrum detected far from the source includes the gravitational redshift from the location of the corona to the detection region).

Some Comptonized photons illuminate the disk and interact with the material on the surface of the disk: Compton scattering and absorption followed by fluorescence emission produce the reflection spectrum. The reflection spectrum in the rest-frame of the material of the disk is characterized by narrow fluorescent emission lines below 10~keV and a Compton hump with a peak around 20-30~keV~\cite{Ross:2005dm,Garcia:2010iz}. The most prominent emission line is usually the iron K$\alpha$ line, which is a very narrow line at 6.4~keV for neutral or weakly ionized iron atoms and shifts up to 6.97~keV for H-like iron ions. However, the reflection spectrum of the whole disk as seen by a distant observer is affected by gravitational redshift and Doppler boosting~\cite{Bambi:2017khi}. The result is that the narrow fluorescent emission lines get broadened. The analysis of these broadened reflection features can be a powerful tool to study the accretion process in the strong gravity region around black holes, measure black hole spins, and test Einstein's theory of General Relativity in the strong field regime~\cite{Bambi:2020jpe}.

\section{Synthetic Spectra}\label{s-ss}

\begin{figure}[t]
\begin{center}
\includegraphics[width=12.0cm]{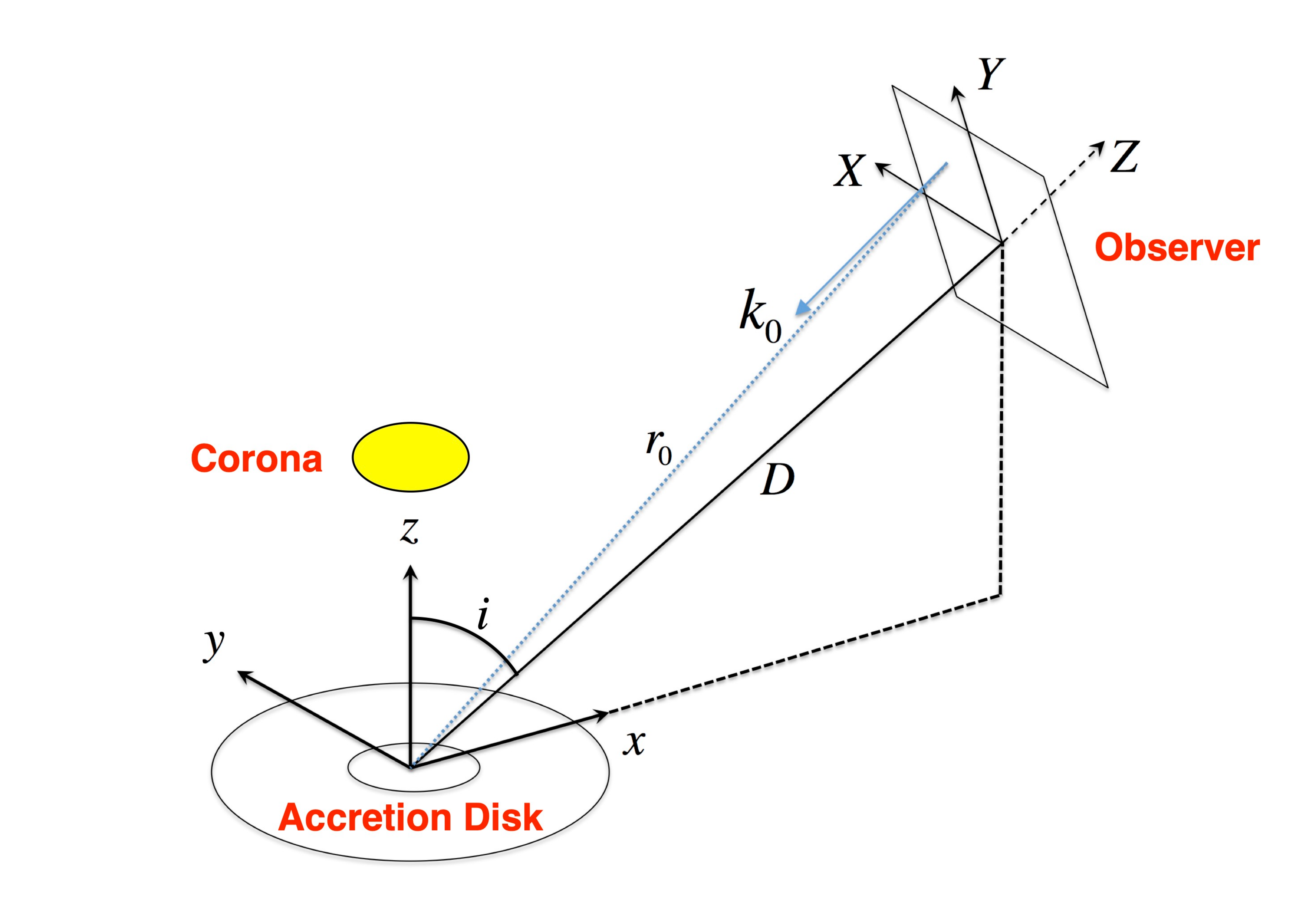}
\end{center}
\vspace{-0.7cm}
\caption{Astrophysical system and distant observer. Figure adapted from Ref.~\cite{Bambi:2016sac} \label{f-setup}}
\end{figure}

If we want to study astrophysical systems like that shown in Fig.~\ref{f-corona}, we have to be able to calculate synthetic spectra, which can then be compared with available X-ray observations in order to measure the properties of the sources. To start, we consider an observer at the distant $D$ from the source and with a viewing angle $i$, as sketched in Fig.~\ref{f-setup}. The image plane of the observer is provided with Cartesian coordinates $(X,Y)$. The calculation of the spectrum of the astrophysical system reduces to the calculation of its image on the image plane of the distant observer. Integrating over the image plane, we find the spectrum of the source as measured by the observer.

The photon flux measured by the observer (for instance, in photons~s$^{-1}$~cm$^{-2}$~keV$^{-1}$) can be written as 
\be\label{eq-flux}
N_{\rm obs} ( E_{\rm obs} ) = \frac{1}{E_{\rm obs}} \int I_{\rm obs} ( E_{\rm obs} , X , Y ) \, d\Omega \, ,
\ee
where $I_{\rm obs}$ is the specific intensity of the radiation on the image plane of the observer, $E_{\rm obs}$ is the photon energy measured by the observer, and the integration is over the solid angle subtended by the image of the astrophysical system on the image plane of the observer. If we can calculate $I_{\rm obs}$, we have the image of the source at any photon energy $E_{\rm obs}$. From Liouville's theorem~\cite{Lindquist:1966igj}, $I/E^3$ is constant along photon paths and therefore we can rewrite Eq.~(\ref{eq-flux}) in terms of the specific intensity of the radiation in the rest-frame of the emitting material $I_{\rm e}$
\be\label{eq-flux2}
N_{\rm obs} ( E_{\rm obs} ) = \frac{1}{E_{\rm obs}} \int g^3 I_{\rm e} ( E_{\rm e} , X , Y ) \, d\Omega \, ,
\ee
where $E_{\rm e}$ is the photon energy at the emission point in the rest-frame of the material and $g = E_{\rm obs}/E_{\rm e}$ is the redshift factor. If we know the properties of the source, we can predict $I_{\rm e}$ and then calculate the observed photon flux $N_{\rm obs}$.

The calculation of the spectrum of the source can proceed as follows. We consider a grid on the image plane of the observer and we fire a photon from every point of the grid to the source in order to calculate its trajectory backwards in time from the detection point on the image plane of the observer to the emission point near the black hole. If the emitting material is optically thick (e.g., the accretion disk), the calculation of the photon trajectory stops when we hit the disk. At that point, we calculate the photon redshift $g$ and the specific intensity of the radiation at the emission point $I_{\rm e}$. If the emitting material is optically thin (e.g., an extended corona), we have to continue the calculation of the photon trajectory and integrate the specific emissivity along the photon path~\cite{Bambi:2013nla}, calculating the redshift $g$ and the specific emissivity at any point.

In the specific case of the system in Fig.~\ref{f-corona}, we have a corona emitting Comptonized photons and an accretion disk emitting a thermal spectrum and a reflection spectrum.

\subsection{Corona} 

If we employ a model with a specific coronal geometry, we can calculate self-consistently the spectrum of the corona. In the rest-frame of the material of the corona, the Comptonized spectrum can be approximated by a power law with a high-energy cutoff\footnote{The shape of the spectrum of photons that inverse Compton scatter off thermal electrons is mainly regulated by the Compton parameter $y = 4 \, (T_{\rm e}/m_{\rm e}) \, {\rm max}(\tau_{\rm e}, \tau_{\rm e}^2)$, where $T_{\rm e}$ is the temperature of the electron gas, $m_{\rm e}$ is the electron mass, and $\tau_{\rm e}$ is the optical depth of the medium. If $y \ll 1$, the spectrum of the Comptonized photons can be approximated by a power law. If $y < 1$, we have a power law with an exponential high-energy cutoff. If $y \gtrsim 1$, the spectrum of the Comptonized photons may not be approximated by a power law with an exponential high-energy cutoff.}, so we have two parameters at every point of the corona: the photon index $\Gamma$ and the high-energy cutoff $E_{\rm cut}$. In the simplest case of the lamppost scenario, the corona is point-like, so there is only one photon index and one high-energy cutoff. The photon index is not affected by the redshift factor between the emission point and the detection point, while the high-energy cutoff of the observed spectrum is $E_{\rm cut, obs} = g E_{\rm cut, e}$, where $E_{\rm cut, e}$ is the high-energy cutoff at the emission point. In the more general case of an extended corona model, calculations should take into account that different parts of the corona may produce Comptonized spectra with different values of $\Gamma$ and $E_{\rm cut}$.

If we do not want to assume a specific coronal geometry, we can simply assume that the total coronal spectrum at the detection point is described by a a power law with photon index $\Gamma$ and high-energy cutoff $E_{\rm cut}$: the available X-ray data of accreting black holes can be fit well with such a simple assumption\footnote{In general, the values of the electron temperature and of the electron density may vary over an extended corona. If most of the Comptonized photons are produced in a relatively homogeneous region of the corona, where it is possible to define an electron temperature and an optical depth, the spectrum of the Comptonized photons can still be approximated well by a power law with an exponential high-energy cutoff.}. In such a case, we can infer the values of the photon index $\Gamma$ and the observed high-energy cutoff $E_{\rm cut}$ from the fit, and we cannot infer the value of the high-energy cutoff at the emission point.

\subsection{Accretion Disk}

For the accretion disk, we normally assume the Novikov-Thorne model~\cite{Novikov:1973kta,Page:1974he}: the disk is on the equatorial plane perpendicular to the black hole spin, the material of the disk follows nearly-geodesic equatorial circular orbits, and the inner edge of the disk is at the innermost stable circular orbit (ISCO) of the spacetime (or at a larger radial coordinate if the disk is truncated). The time-averaged radial structure of the disk is completely determined by imposing the conservation of mass, energy, and angular momentum. The time-averaged energy flux emitted from the disk surface at the radial coordinate $r$ is
\be\label{eq-F(r)}
\mathcal{F} (r) = \frac{\dot{M}}{4 \pi M^2} F(r) \, ,
\ee
where $M$ is the black hole mass, $\dot{M}$ is the black hole mass accretion rate, and $F(r)$ is a dimensionless function whose expression can be found in Refs.~\cite{Bambi:2017khi,Page:1974he}. The effective temperature of the disk $T_{\rm eff} (r)$ can be defined from the Stefan-Boltzmann law $\mathcal{F} = \sigma T_{\rm eff}^4$, where $\sigma$ is the Stefan-Boltzmann constant. The interactions of thermal photons with the hot atmosphere above the accretion disk can be taken into account simply introducing a (dimensionless) hardening factor $f_{\rm col}$, which is normally thought to be in the range 1.5 to 1.9 for stellar-mass black holes~\cite{Shimura:1995nu}. The color temperature of the disk is $T_{\rm col} = f_{\rm col} T_{\rm eff}$ and the local specific intensity of the radiation on the disk surface is
\be\label{eq-bb}
I_{\rm e} ( E_{\rm e} ) = \frac{E_{\rm e}^3}{2 \pi^2} \frac{1}{f_{\rm col}^4} \frac{\Upsilon}{\exp\left( E_{\rm e} / T_{\rm col}\right) - 1} \, ,
\ee 
where $\Upsilon = \Upsilon (\vartheta_{\rm e})$ is, in general, a function of the angle between the photon 4-momentum and the normal of the disk surface, $\vartheta_{\rm e}$. $\Upsilon = 1$ for isotropic emission and $\Upsilon = 0.5 + 0.75 \cos\vartheta_{\rm e}$ for limb-darkened emission.

In the case of the reflection spectrum, one has to consider the illumination of some material by radiation with a certain spectrum (e.g., a power law with a high-energy cutoff) and incident angle $\vartheta_{\rm i}$. The calculations require to solve radiative transfer equations and determine the spectrum of the reflection radiation at the emission angle $\vartheta_{\rm e}$. The details can be found in Refs.~\cite{Ross:2005dm,Garcia:2010iz}. Current models employ a number of simplifications. For example, the electron density of the disk is assumed to be constant along the vertical direction\footnote{Non-relativistic reflection spectra in which the electron density is determined from hydrostatic balance were studied in Ref.~\cite{Nayakshin:2000vm}, where the authors found qualitatively and quantitatively different results from those obtained using the constant density assumption. A new study on the impact of the vertical profile of the electron density on non-relativistic and relativistic reflection spectra should be reported soon in Huang et al. (in preparation), and we plan to implement the possibility of a non-constant electron density profile in the disk's vertical direction in {\tt relxill\_nk} in the future.}. In ${\tt reflionx}$, the output reflection spectrum is averaged over the emission angle $\vartheta_{\rm e}$~\cite{Ross:2005dm}. In ${\tt xillver}$, the output reflection spectrum is calculated for a grid of emission angles $\vartheta_{\rm e}$, but the incident radiation is always assumed to illuminate the disk with an inclination $\vartheta_{\rm i} = 45$~deg~\cite{Garcia:2010iz}\footnote{The calculation of the correct incident angle $\vartheta_{\rm i}$ at every point of the disk would be possible only for a specific coronal geometry.}.

\subsection{Transfer Function of the Disk}

\begin{figure}[t]
\begin{center}
\includegraphics[width=10.0cm]{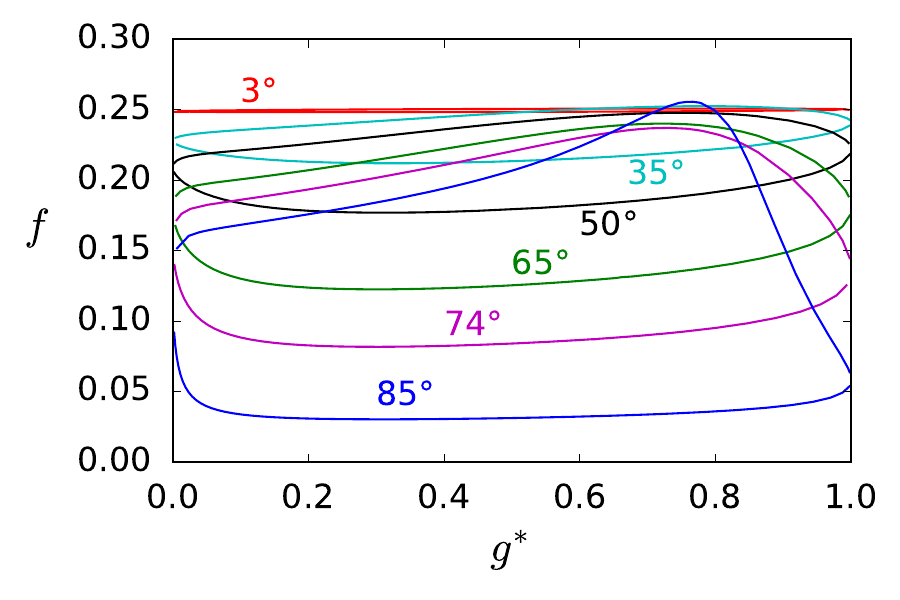}
\end{center}
\vspace{-0.7cm}
\caption{Transfer function of an infinitesimally-thin Novikov-Thorne disk in the Kerr spacetime for spin parameter $a_* = 0.998$, emission radius $r_{\rm e} = 4$~$r_{\rm g}$ ($r_{\rm g} = M$ is the gravitational radius), and different values of the viewing angle of the observer $i$ ranging from 3~deg to 85~deg. Figure from Ref.~\cite{Bambi:2016sac} \label{f-trf}}
\end{figure}

The calculations described above for the disk spectra turn out to be very time consuming. They can be done to calculate a single spectrum with certain input parameters, but they are not suitable to calculate quickly many spectra during the data analysis process, when we want to compare a large number of synthetic spectra obtained from different values of the input parameters with the observed data in order to find the best-fit and a measurement for all the parameters of the model. To solve this problem, the current strategy is to rewrite Eq.~(\ref{eq-flux2}) with the transfer function~\cite{Bambi:2017khi,Cunningham:1975zz}
\be\label{eq-trf}
N_{\rm obs} ( E_{\rm obs} ) = \frac{1}{E_{\rm obs}} \frac{1}{D^2} \int_{r_{\rm in}}^{r_{\rm out}} dr_{\rm e} \int_0^1 dg^* \,
\frac{\pi r_{\rm e} g^2}{\sqrt{g^* ( 1 - g^* )}} \; f (g^* , r_{\rm e} , i ) \; I_{\rm e} ( E_{\rm e} , r_{\rm e} , \vartheta_{\rm e} ) \, ,
\ee  
where $r_{\rm e}$ is the radial coordinate of the emission point on the disk, $r_{\rm in}$ and $r_{\rm out}$ are, respectively, the inner and the outer edges of the disk, $g^*$ is the relative redshift factor
\be
g^* = \frac{g - g_{\rm min}}{g_{\rm max} - g_{\rm min}} \, , 
\ee
$g_{\rm max} = g_{\rm max} (r_{\rm e},i)$ and $g_{\rm min} = g_{\rm min} (r_{\rm e},i)$ are, respectively, the maximum and the minimum values of the redshift factor $g$ for the photons emitted from the radial coordinate $r_{\rm e}$ and detected by a distant observer with viewing angle $i$, $f$ is the transfer function
\be
f (g^* , r_{\rm e} , i ) = \frac{g \sqrt{g^* \left(1 - g^*\right)}}{\pi r_{\rm e}} 
\left| \frac{\partial \left( X , Y \right)}{\partial \left( g^* , r_{\rm e}\right)} \right| \, ,
\ee
and $|\partial \left( X , Y \right) / \left( g^* , r_{\rm e}\right)|$ is the Jacobian between the coordinates of the image plane of the observer $(X,Y)$ and the coordinates on the disk $(g^* , r_{\rm e})$. The transfer function depends on the spacetime metric (e.g., the Kerr metric), the accretion disk model (e.g., the Novikov-Thorne model), and is a function of $g^*$ and $r_{\rm e}$ for a given viewing angle $i$. In the Kerr spacetime and for an infinitesimally thin disk, the transfer function at any emission radius $r_{\rm e}$ is a closed curve as shown in Fig.~\ref{f-trf}. In the Kerr spacetime but in the presence of a disk of finite thickness, the curve may not be closed for some range of the emission radius $r_{\rm e}$~\cite{Abdikamalov:2020oci}.

With the expression in Eq.~(\ref{eq-trf}), we can calculate quickly many spectra. In the case of the thermal spectrum, we can tabulate the transfer function $f (g^* , r_{\rm e} , i )$ and the dimensionless function $F (r)$ into FITS files\footnote{FITS (Flexible Image Transport System) is the data format most used in astronomy for storing data files.}. For example, let us assume that the spacetime is described by the Kerr solution and the disk is described by an infinitesimally-thin Novikov-Thorne disk. The transfer function can be tabulated for a grid of different values of black hole spins $a_*$ and viewing angles of the observer $i$. For every pair $(a_*,i)$, the transfer function can be evaluated for a grid of emission radii $r_{\rm e}$ and relative redshift factors $g^*$. The function $F(r)$ can be evaluated for a grid of black hole spins $a_*$ and emission radii $r_{\rm e}$. The integral in Eq.~(\ref{eq-trf}) is now straightforward to calculate: there are no ray-tracing calculations to do and the model calls the FITS files in which $f (g^* , r_{\rm e} , i )$ and $F (r)$ are tabulated to calculate the integral. In the case of the reflection spectrum, we can tabulate the transfer function $f (g^* , r_{\rm e} , i )$ and the reflection spectrum in the rest-frame of the material. If we assume a specific coronal geometry, we can also tabulate the emissivity profile of the reflection component in a grid for the parameters of the corona. As in the case of the thermal spectrum, the model does not have to solve time consuming equations: it calls the FITS files in which the transfer function, the reflection spectrum, and the possible emissivity profile are tabulated and performs the integral to determine the observed spectrum.


\section{${\tt ABHModels}$}\label{s-models}

Currently we have the following six public models on ${\tt ABHModels}$\footnote{${\tt ABHModels}$ is the acronym for Accreting Black Hole Models. The URL is \href{https://github.com/ABHModels}{https://github.com/ABHModels}}:

1. ${\tt blackray}$

2. ${\tt nkbb}$

3. ${\tt relxill\_nk}$

4. ${\tt raytransfer}$

5. ${\tt F}$-${\tt code}$

6. ${\tt blacklamp}$

The next subsections will provide a description for every model. For support and inquiries, you can contact us at \href{mailto:relxill_nk@fudan.edu.cn}{relxill\_nk@fudan.edu.cn}.

\subsection{${\tt blackray}$}\label{ss-blackray}

${\tt blackray}$\footnote{\href{https://github.com/ABHModels/blackray}{https://github.com/ABHModels/blackray}} is a ray-tracing code to calculate spectra of infinitesimally-thin Novikov-Thorne disks around compact objects in stationary, axisymmetric, and asymptotically-flat spacetimes. In the current public version, the code calculates iron line profiles and full reflection spectra of infinitesimally-thin Novikov-Thorne disks in the Johannsen spacetime with deformation parameters $\alpha_{13}$, $\alpha_{22}$, $\alpha_{52}$, and $\epsilon_3$ (higher order deformation parameters are assumed to vanish)~\cite{Johannsen:2013szh}. The Kerr solution is recovered for $\alpha_{13} = \alpha_{22} = \alpha_{52} = \epsilon_3 = 0$.

The model requires a ${\tt xillver}$ table\footnote{The latest ${\tt xillver}$ tables can be download from 

\href{http://www.sternwarte.uni-erlangen.de/~dauser/research/relxill/index.html}{http://www.sternwarte.uni-erlangen.de/~dauser/research/relxill/index.html}}. To run the code, there is the Python script ${\tt run.py}$. Before running the Python script, it is necessary to specify the location of the ${\tt xillver}$ table (line~9 in ${\tt run.py}$) as well as all the parameters of the model (lines~10-23 in ${\tt run.py}$). The parameters {\tt rstep} and {\tt pstep} regulate the resolution of the image of the observer and, in turn, the number of photons fired to the disk. In general, we suggest the following values 

\noindent {\tt rstep} = 1.008

\noindent {\tt pstep} = 2$\pi$/720

\noindent In some special cases (for example, to calculate synthetic spectra to use with the next generation of X-ray missions), we have to increase the resolution and we may need {\tt rstep} = 1.0001 and {\tt pstep} = 2$\pi$/3600 (see Ref.~\cite{Abdikamalov:2013}). To run the Python script, the command is

\noindent {\tt python3 run.py}

\noindent ${\tt run.py}$ generates two outputs: an output for a relativistically broadened iron line profile and another output for a relativistically broadened full reflection spectrum.

\subsubsection*{Notes}

In the Johannsen spacetime, equatorial circular orbits are always vertically stable and the ISCO radius (which is assumed to set the inner edge of the accretion disk in ${\tt blackray}$) is determined by the stability along the radial direction. In the current version of the code, the ISCO radius is thus calculated checking the stability along the radial direction only. However, there are non-Kerr spacetimes in which the ISCO radius is determined by the stability of the orbit along the vertical direction; see, e.g., Refs.~\cite{Bambi:2011vc,Bambi:2013eb}. If you change the metric and the new metric can have equatorial circular orbits that are vertically unstable, it is necessary to modify even the subroutine to determine the ISCO radius.

\subsection{${\tt nkbb}$}\label{ss-nkbb}

${\tt nkbb}$\footnote{\href{https://github.com/ABHModels/nkbb}{https://github.com/ABHModels/nkbb}} is an ${\tt XSPEC}$ model for thermal spectra of infinitesimally-thin Novikov-Thorne disks around compact objects in stationary, axisymmetric, and asymptotically-flat spacetimes~\cite{Zhou:2019fcg}. We have also implemented the option to use ${\tt nkbb}$ to calculate thermal spectra of Novikov-Thorne disks of finite thickness around Kerr black holes~\cite{Zhou:2020koa}.

\subsubsection*{Parameters of the model}

The model has 12 parameters:

\textcolor{blue}{${\tt eta}$}: It is a parameter to regulate the location of the inner edge of the accretion disk. It is defined by the relation $R_{\rm in} = \left( 1 + {\tt eta} \right) R_{\rm ISCO}$, where $R_{\rm in}$ is the radial coordinate of the inner edge of the accretion disk and $R_{\rm ISCO}$ is the radial coordinate of the ISCO. While in most situations one assumes ${\tt eta} = 0$, for some observations one may want to consider the possibility of a truncated disk with $R_{\rm in} > R_{\rm ISCO}$

\textcolor{blue}{${\tt a*}$}: Spin parameter of the compact object.

\textcolor{blue}{${\tt i}$}: Inclination angle of the disk with respect to the line of sight of the observer. It is measured in deg.

\textcolor{blue}{${\tt Mbh}$}: Mass of the compact object. It is measured in Solar Mass units.

\textcolor{blue}{${\tt Mdd}$}: Mass accretion rate. It is measured in units of $10^{18}$~g~s$^{-1}$.

\textcolor{blue}{${\tt Dbh}$}: Distance of the source from the observer. It is measured in kpc.

\textcolor{blue}{${\tt hd}$}: Hardening factor, also known as the color-correction factor; it is $f_{\rm col}$ in Eq.~(\ref{eq-bb}). 

\textcolor{blue}{${\tt rflag}$}: It is a flag to switch on/off the effect of returning radiation. However, in the current public version of ${\tt nkbb}$ this function is not implemented, so the value of ${\tt rflag}$ has no impact on the output spectra.

\textcolor{blue}{${\tt lflag}$}: It is a flag to switch on/off the effect of limb-darkening. If ${\tt lflag} \le 0$, the emission is isotropic; $\Upsilon = 1$ in Eq.~(\ref{eq-bb}). If ${\tt lflag} > 0$, the emission is limb-darkened; $\Upsilon = 0.5 + 0.75 \cos\vartheta_{\rm e}$ in Eq.~(\ref{eq-bb}). This parameter should never be free. 

\textcolor{blue}{${\tt defpar\_type}$}: This {\it integer} parameter is used to regulate the geometry of the spacetime.   ${\tt defpar\_type}=0$ corresponds to the Kerr spacetime. In such a case, ${\tt defpar\_value}$ is used to regulate the mass accretion rate and ranges from 0 to 1, corresponding to the Eddington-scaled mass accretion rate 0 (infinitesimally-thin disk) to 0.3 (see Ref.~\cite{Zhou:2020koa} for more details). ${\tt defpar\_type}=1$ corresponds to the Johannsen spacetime with non-vanishing deformation parameter $\alpha_{13}$ (all other deformation parameters vanish). ${\tt defpar\_type}=2$ corresponds to the Johannsen spacetime with non-vanishing deformation parameter $\alpha_{22}$ (all other deformation parameters vanish). ${\tt defpar\_type}=3$ corresponds to the Johannsen spacetime with non-vanishing deformation parameter $\epsilon_{3}$ (all other deformation parameters vanish). For ${\tt defpar\_type}=1,2,3$, ${\tt defpar\_value}$ ranges from $-1$ to 1 but it is not the value of the deformation parameter appearing in the Johannsen metric. It is the {\it scaled} value of the deformation parameter (see below).

\textcolor{blue}{${\tt defpar\_value}$}: If ${\tt defpar\_type}=0$, ${\tt defpar\_value}$ corresponds to the scaled value of the mass accretion rate, which regulates the thickness of the accretion disk. If ${\tt defpar\_type}=1,2,3$, ${\tt defpar\_value}$ corresponds to the scaled value of the deformation parameter. To obtain the Eddington-scaled mass accretion rate (${\tt defpar\_type}=0$) and the unscaled deformation parameter appearing in the Johannsen metric (${\tt defpar\_type}=1,2,3$), one can use the Python scripts ${\tt unscale.py}$ and ${\tt unscale\_batch.py}$ (see below).

\textcolor{blue}{${\tt norm}$}: This is the normalization of the component and should be set to 1 because the luminosity of the source is not arbitrary and is determined by the value of the parameters of the model.

\subsubsection*{How to use the model}

${\tt nkbb}$ requires a FITS file for the transfer functions (which can be generated by ${\tt raytransfer}$) and a FITS file for the function $F(r)$ (which can be generated by ${\tt F}$-${\tt code}$). The FITS files are also available upon request by contacting us at \href{mailto:relxill_nk@fudan.edu.cn}{relxill\_nk@fudan.edu.cn}. Please note that the FITS file of the transfer functions of ${\tt nkbb}$ is not the FITS file of the transfer functions of ${\tt relxill\_nk}$ because the transfer functions for ${\tt nkbb}$ must be tabulated over larger accretion disks (see Subsection~\ref{ss-rt}).

To use ${\tt nkbb}$ with ${\tt XSPEC}$~\cite{xspec}, first, you have to open the file ${\tt nkbb.h}$ and change the variable TR\_TABLE\_PATH (line~22) to the path of the FITS files. The names of the FITS files of the transfer functions are at lines~25-31 and the names of the FITS files of the functions $F(r)$ are at lines~33-36. Second, you have to create the model with the script ${\tt compile\_NKBB.sh}$. The line commands are:

\noindent {\tt chmod u+r compile\_NKBB.sh}

\noindent {\tt ./compile\_NKBB.sh}

\noindent Last, it is necessary to load the model in ${\tt XSPEC}$. If you open ${\tt XSPEC}$ in the directory of the model, it is enough the following line command

\noindent {\tt lmod nkbb .}

\noindent Otherwise, you need to specify the the path of the model

\noindent {\tt lmod nkbb [path of the model]}

As mentioned before, ${\tt defpar\_value}$ corresponds to the scaled deformation parameter. To calculate the unscaled deformation parameter appearing in the metric, there are two Python scripts: ${\tt unscale.py}$ and ${\tt unscale\_batch.py}$.

${\tt unscale.py}$ is to unscale a single value. The command line is

\noindent {\tt python3 unscale.py [defpar\_type] [a*] [defpar\_value]}

\noindent where {\tt [defpar\_type]} is the value of ${\tt defpar\_type}$ (${\tt defpar\_type} = 0, 1, 2, 3$), {\tt [a*]} is the value of the black hole spin parameter, and {\tt [defpar\_value]} is the value of the scaled parameter. The output is the value of the unscaled deformation parameter. The Python script reads the FITS file and calculates the unscaled value from the scaled one. If you change FITS file or the FITS file is not in the working directory, you have to change lines~7-24 in ${\tt unscale.py}$.

If you scan the spin parameter vs deformation parameter plane with the ${\tt steppar}$ command of ${\tt XSPEC}$ and you generate a plain .txt file, you can unscale the deformation parameter with ${\tt unscale\_batch.py}$. The command line is

\noindent {\tt python3 unscale\_batch.py [defpar\_type] [input file] [output file]}

\noindent where {\tt [input file]} is the plain file from ${\tt XSPEC}$ and {\tt [output file]} is the output file of the Python script and can be plotted to show different contour levels on the plane spin parameter vs deformation parameter. If you change FITS file or the FITS file is not in the working directory, you have to change lines~12-29 of ${\tt unscale\_batch.py}$.

\subsection{${\tt relxill\_nk}$}

${\tt relxill\_nk}$\footnote{\href{https://github.com/ABHModels/relxill_nk}{https://github.com/ABHModels/relxill\_nk}} is an ${\tt XSPEC}$ model for reflection spectra of Novikov-Thorne disks around compact objects in stationary, axisymmetric, and asymptotically-flat spacetimes~\cite{Bambi:2016sac,Abdikamalov:2019yrr,Abdikamalov:2020oci,Riaz:2020svt,Abdikamalov:2021rty,Abdikamalov:2021ues}. It is an extension of the ${\tt relxill}$ package developed by Thomas Dauser and Javier Garcia~\cite{Dauser:2013xv,Garcia:2013oma,Garcia:2013lxa} to non-Kerr spacetimes (${\tt nk}$ at the end of the its name stands for Non-Kerr). Note that the current version of ${\tt relxill\_nk}$ does not use the latest {\tt xillver} table, but the previous version in which {\tt xillverD} and {\tt xillverCp} are two different models.

\subsubsection*{Flavors}

In the current version, there are 22 {\it flavors}:

\textcolor{magenta}{${\tt relline\_nk}$}: Model for relativistically broadened line profiles (i.e., the spectrum at the emission point in the rest-frame of the material in the disk is a narrow line) from Novikov-Thorne disks around compact objects in stationary, axisymmetric, and asymptotically-flat spacetimes~\cite{Bambi:2016sac}. By default, the non-relativistic line is at 6.4~keV. The thickness of the disk can be finite and regulated by the mass accretion rate~\cite{Abdikamalov:2020oci}. The emissivity profile of the accretion disk is described by a twice broken power law (5~parameters: emissivity index of the inner region, ${\tt Index1}$; emissivity index of the central region, ${\tt Index2}$; emissivity index of the outer region, ${\tt Index3}$; breaking radius between the inner and central regions, ${\tt Rbr1}$; breaking radius between the central and outer regions, ${\tt Rbr2}$). 

\textcolor{magenta}{${\tt rellinelp\_nk}$}: Model for relativistically broadened line profiles~\cite{Abdikamalov:2019yrr}. It is like ${\tt relline\_nk}$, but the emissivity profile is that of the lamppost model and it is fully regulated by the height of the corona, ${\tt h}$ (in units of gravitational radii).

\textcolor{magenta}{${\tt relconv\_nk}$}: Convolution model for calculating spectra of Novikov-Thorne disks around compact objects in stationary, axisymmetric, and asymptotically-flat spacetimes~\cite{Bambi:2016sac}. If the spectrum at the emission point in the rest-frame of the material in the disk is described by the ${\tt XSPEC}$ model ${\tt spectrum}$, the relativistic spectrum observed far from the source is 

\noindent ${\tt relconv\_nk}\times{\tt spectrum}$.

\noindent If ${\tt spectrum}$ is a narrow line, the output is the same as ${\tt relline\_nk}$. The emissivity profile of the accretion disk is described by a twice broken power law.

\textcolor{magenta}{${\tt relconvlp\_nk}$}: Convolution model for calculating the spectra of Novikov-Thorne disks~\cite{Abdikamalov:2019yrr}. It is like ${\tt relconv\_nk}$, but the emissivity profile is that of the lamppost model and it is fully regulated by the height of the corona, ${\tt h}$ (in units of gravitational radii).

\textcolor{magenta}{${\tt xillver}$}: Model for non-relativistic reflection spectra~\cite{Garcia:2013oma}. The spectrum of the illuminating radiation is described by a power law with a high-energy cutoff (2~parameters: photon index ${\tt gamma}$ and high-energy cutoff ${\tt Ecut}$). The disk electron density is set to $10^{15}$~cm$^{-3}$. This is the ${\tt xillver}$ model of the ${\tt relxill}$ package developed by Thomas Dauser and Javier Garcia; see Ref.~\cite{Garcia:2013oma}.

\textcolor{magenta}{${\tt xillverD}$}: Model for non-relativistic reflection spectra~\cite{Garcia:2013oma}. It is like ${\tt xillver}$, but the disk electron density can be allowed to vary from $10^{15}$~cm$^{-3}$ to $10^{19}$~cm$^{-3}$ or more (depending on the version of the FITS file).

\textcolor{magenta}{${\tt xillverCp}$}: Model for non-relativistic reflection spectra~\cite{Garcia:2013oma}. It is like ${\tt xillver}$, but the illuminating radiation is described by the ${\tt nthcomp}$ model~\cite{Zdziarski:1996wq,Zdziarski:1998} and it is regulated by the photon index ${\tt gamma}$ and the coronal temperature ${\tt kTe}$ (in keV).

\textcolor{magenta}{${\tt relxill\_nk}$}: Model for relativistically broadened reflection spectra of Novikov-Thorne disks around compact objects in stationary, axisymmetric, and asymptotically-flat spacetimes~\cite{Bambi:2016sac}. It employs the non-relativistic reflection spectra of ${\tt xillver}$. The thickness of the disk can be finite and regulated by the mass accretion rate~\cite{Abdikamalov:2020oci}. The emissivity profile of the accretion disk is described by a twice broken power law (5~parameters: emissivity index of the inner region, ${\tt Index1}$; emissivity index of the central region, ${\tt Index2}$; emissivity index of the outer region, ${\tt Index3}$; breaking radius between the inner and central regions, ${\tt Rbr1}$; breaking radius between the central and outer regions, ${\tt Rbr2}$). Note that it is not exactly like 

\noindent ${\tt relconv\_nk}\times{\tt xillver}$

\noindent because the non-relativistic reflection spectra in ${\tt xillver}$ depend on the emission angle $\vartheta_{\rm e}$: ${\tt relxill\_nk}$ uses the actual emission angles from the ${\tt xillver}$ table while ${\tt relconv\_nk}$ uses the emission angle equal to the viewing angle of the observer $i$.

\textcolor{magenta}{${\tt relxilllp\_nk}$}: Model for relativistically broadened reflection spectra~\cite{Abdikamalov:2019yrr}. It is like ${\tt relxill\_nk}$, but the emissivity profile is that of the lamppost model and it is fully regulated by the height of the corona, ${\tt h}$ (in units of gravitational radii).

\textcolor{magenta}{${\tt relxillD\_nk}$}: Model for relativistically broadened reflection spectra~\cite{Abdikamalov:2019yrr}. It is like ${\tt relxill\_nk}$, but it employs the non-relativistic reflection spectra of ${\tt xillverD}$, so the electron density ranges from $10^{15}$~cm$^{-3}$ to $10^{19}$~cm$^{-3}$ or more (depending on the version of the FITS file).

\textcolor{magenta}{${\tt relxilllpD\_nk}$}: Model for relativistically broadened reflection spectra~\cite{Abdikamalov:2019yrr}. It employs the lamppost corona model and the non-relativistic reflection spectra of ${\tt xillverD}$.

\textcolor{magenta}{${\tt relxillCp\_nk}$}: Model for relativistically broadened reflection spectra~\cite{Abdikamalov:2019yrr}. It is like ${\tt relxill\_nk}$, but it employs the non-relativistic reflection spectra of ${\tt xillverCp}$ and the continuum from the corona is described by ${\tt nthcomp}$.

\textcolor{magenta}{${\tt relxilllpCp\_nk}$}: Model for relativistically broadened reflection spectra~\cite{Abdikamalov:2019yrr}. It employs the lamppost corona model and the non-relativistic reflection spectra of ${\tt xillverCp}$. The continuum from the corona is described by ${\tt nthcomp}$.

\textcolor{magenta}{${\tt relxillion\_nk}$}: Model for relativistically broadened reflection spectra~\cite{Abdikamalov:2021rty}. It is like ${\tt relxill\_nk}$, but the ionization parameter profile is described by a power law. There are 2~parameters: the logarithm of the ionization parameter at the inner edge of the accretion disk, ${\tt logxi}$, and the ionization index, ${\tt xi\_index}$. The ionization parameter at the radial coordinate $r$ is given by
\be
\xi (r) = \xi_{\rm in} \left(\frac{R_{\rm in}}{r}\right)^{\alpha_\xi} \, ,
\ee
where $\xi_{\rm in}$ is the ionization parameter at the inner edge of the accretion disk, $R_{\rm in}$ is the inner edge of the accretion disk, and $\alpha_\xi$ is ${\tt xi\_index}$.

\textcolor{magenta}{${\tt relxillionCp\_nk}$}: Model for relativistically broadened reflection spectra~\cite{Abdikamalov:2021rty}. It is like ${\tt relxillion\_nk}$, but it employs the non-relativistic reflection spectra of ${\tt xillverCp}$ and the continuum from the corona is described by ${\tt nthcomp}$.

\textcolor{magenta}{${\tt relxilllpion\_nk}$}: Model for relativistically broadened reflection spectra~\cite{Abdikamalov:2021rty}. It is like ${\tt relxillion\_nk}$, but the emissivity profile is that of the lamppost model and it is fully regulated by the height of the corona, ${\tt h}$ (in units of gravitational radii).

\textcolor{magenta}{${\tt relxilllpionCp\_nk}$}: Model for relativistically broadened reflection spectra~\cite{Abdikamalov:2021rty}. It is like ${\tt relxillion\_nk}$, but the emissivity profile is that of the lamppost corona, the non-relativistic reflection spectra are those of ${\tt xillverCp}$, and the continuum from the corona is described by ${\tt nthcomp}$.

\textcolor{magenta}{${\tt relxilldgrad\_nk}$}: Model for relativistically broadened reflection spectra~\cite{Abdikamalov:2021ues}. It is like ${\tt relxill\_nk}$, but the electron density profile is described by a power law and the ionization profile is calculated accordingly. There are 3~parameters: the logarithm of the electron density at the inner edge of the accretion disk, ${\tt logN}$, the electron density index, ${\tt logN\_index}$, and the logarithm of the maximum value of the ionization parameter, ${\tt logxi}$. The electron density at the radial coordinate $r$ is given by
\be
n (r) = n_{\rm in} \left(\frac{R_{\rm in}}{r}\right)^{\alpha_n} \, ,  
\ee
where $n_{\rm in}$ is the electron density at the inner edge of the accretion disk, $R_{\rm in}$ is the inner edge of the accretion disk, and $\alpha_n$ is ${\tt logN\_index}$. The ionization parameter at the radial coordinate $r$ is given by
\be
\xi (r) = \xi_{\rm max} \left[\frac{4 \, \pi \, \epsilon(r)}{n(r)}\right]_{\rm Norm} \, ,
\ee
where $\xi_{\rm max}$ is ${\tt logxi}$ (which is not necessarily the ionization parameter at the inner edge of the accretion disk), $\epsilon(r)$ is the emissivity profile at the radial coordinate $r$, and Norm is used to indicate that the expression in square brackets is normalized with respect to its maximum value reached on the accretion disk.

\textcolor{magenta}{${\tt relxillring\_nk}$}: Model for relativistically broadened reflection spectra~\cite{Riaz:2020svt}. It is like ${\tt relxill\_nk}$, but the emissivity profile is that generated by a ring-like corona, which is regulated by the height, ${\tt h}$, and the radius, ${\tt r\_ring}$, of the corona (both in units of gravitational radii).

\textcolor{magenta}{${\tt relxilldisk\_nk}$}: Model for relativistically broadened reflection spectra~\cite{Riaz:2020svt}. It is like ${\tt relxill\_nk}$, but the emissivity profile is that generated by a disk-like corona, which is regulated by the height, ${\tt h}$, the inner radius, ${\tt rdisk\_in}$, and outer radius, ${\tt rdisk\_out}$, of the corona (all in units of gravitational radii). 

\textcolor{magenta}{${\tt rellinering\_nk}$}: Model for relativistically broadened line profiles~\cite{Riaz:2020svt}. It is like ${\tt relline\_nk}$, but the emissivity profile is that generated by a ring-like corona, which is regulated by the height, ${\tt h}$, and the radius, ${\tt r\_ring}$, of the corona (both in units of gravitational radii).

\textcolor{magenta}{${\tt rellinedisk\_nk}$}: Model for relativistically broadened line profiles~\cite{Riaz:2020svt}. It is like ${\tt relline\_nk}$, but the emissivity profile is that of generated by disk-like corona, which is regulated by the height, ${\tt h}$, the inner radius, ${\tt rdisk\_in}$, and outer radius, ${\tt rdisk\_out}$, of the corona (all in units of gravitational radii).

\subsubsection*{Parameters of the model}

The model has the following parameters (but every flavor has only some of these parameters, as one can easily see opening the file ${\tt lmodel\_relxill.dat}$):

\textcolor{blue}{${\tt lineE}$}: This parameter appears in the models for relativistically broadened line profiles and corresponds to the energy of the line in the rest frame of the material in the disk. It is in keV and its default value is 6.4~keV, which corresponds to the energy of the Fe~K$\alpha$ line for neutral iron. 

\textcolor{blue}{${\tt Index1}$}: Emissivity index of the inner part of the accretion disk. The emissivity profile is $\epsilon(r) \propto 1/r^{q_1}$ for $r < r_1$, where $q_1 = {\tt Index1}$ and $r_1 = {\tt Rbr1}$. 

\textcolor{blue}{${\tt Index2}$}: Emissivity index of the central part of the accretion disk. The emissivity profile is $\epsilon(r) \propto 1/r^{q_2}$ for $r_1 < r < r_2$, where $q_2 = {\tt Index2}$, $r_1 = {\tt Rbr1}$, and $r_2 = {\tt Rbr2}$.

\textcolor{blue}{${\tt Index3}$}: Emissivity index of the outer part of the accretion disk. The emissivity profile is $\epsilon(r) \propto 1/r^{q_3}$ for $r > r_2$, where $q_3 = {\tt Index3}$ and $r_2 = {\tt Rbr2}$.

\textcolor{blue}{${\tt Rbr1}$}: Radial coordinate between the inner and central parts of the disk. It is in units of gravitational radii.

\textcolor{blue}{${\tt Rbr2}$}: Radial coordinate between the central and outer parts of the disk. It is in units of gravitational radii.

\textcolor{blue}{${\tt a}$}: Spin parameter of the compact object.

\textcolor{blue}{${\tt Incl}$}: Inclination angle of the disk with respect to the line of sight of the observer. It is in degree.

\textcolor{blue}{${\tt Rin}$}: Radial coordinate of the inner edge of the accretion disk. If positive, it is in units of gravitational radii. If negative, it is in units of the ISCO radius. To set the inner edge of the accretion disk at the ISCO radius, it is enough to freeze ${\tt Rin}$ to $-1$.

\textcolor{blue}{${\tt Rout}$}: Radial coordinate of the outer edge of the accretion disk. Normally it is set to a value large enough that its exact value does not matter, as the emission at large radii is negligible.

\textcolor{blue}{${\tt z}$}: Redshift of the source. For Galactic sources, it can be set to 0, as typical velocities are at the level of 200~km~s$^{-1}$ and can be ignored. For extragalactic sources, it may be set to the redshift of the source, which is usually known from other measurements.

\textcolor{blue}{${\tt limb}$}: This parameter appears in the models for relativistically broadened line profiles and is a flag to select the emission type. If ${\tt lflag} = 0$, the emission is isotropic; if ${\tt lflag} = 1$, the emission is limb-darkened; if ${\tt lflag} = 2$, the emission is limb-brightened.

\textcolor{blue}{${\tt def\_par\_type}$}: This {\it integer} parameter is used to regulate the geometry of the spacetime. ${\tt def\_par\_type}=1$ corresponds to the Johannsen spacetime with non-vanishing deformation parameter $\alpha_{13}$ (all other deformation parameters vanish). ${\tt def\_par\_type}=2$ corresponds to the Johannsen spacetime with non-vanishing deformation parameter $\alpha_{22}$ (all other deformation parameters vanish). ${\tt def\_par\_type}=3$ corresponds to the Johannsen spacetime with non-vanishing deformation parameter $\epsilon_{3}$ (all other deformation parameters vanish). Higher values of ${\tt def\_par\_type}$ corresponds to other black hole spacetimes (the names of the FITS files and the spacetimes can be found in ${\tt relbase.h}$).

\textcolor{blue}{${\tt def\_par\_value}$}: It is the scaled value of the deformation parameter and usually ranges from $-1$ to 1 (but for some spacetimes it may ranges from 0 to 1). To obtain the unscaled deformation parameter appearing in the metric, one can use the Python scripts ${\tt unscale.py}$ and ${\tt unscale\_batch.py}$, as discussed in Subsection~\ref{ss-nkbb}.

\textcolor{blue}{${\tt mdot\_type}$}: This {\it integer} parameter is used to regulate the thickness of the accretion disk~\cite{Abdikamalov:2020oci}. If ${\tt mdot\_type} = 0$, the disk is infinitesimally-thin. If ${\tt mdot\_type} = 1$ (respectively 2, 3 and 4), the Eddington-scaled mass accretion rate is 5\% (respectively 10\%, 20\%, and 30\%). 

\textcolor{blue}{${\tt h}$}: Height of the lamppost corona in units of gravitational radii. This parameter appears in the models that assume a lamppost corona.

\textcolor{blue}{${\tt gamma}$}: Photon index of the radiation illuminating the disk. 

\textcolor{blue}{${\tt Ecut}$}: High-energy cutoff of the radiation illuminating the disk in the rest-frame of the observer. It is given in keV. 

\textcolor{blue}{${\tt logxi}$}: Logarithm of the ionization parameter, where the ionization parameter is expressed in erg~cm~s$^{-1}$. In the models in which the ionization parameter profile is described by a power law, it corresponds to the logarithm of the ionization parameter at the inner edge of the disk~\cite{Abdikamalov:2021rty}. In ${\tt relxilldgrad\_nk}$, it corresponds to the maximum value of the logarithm of the ionization parameter~\cite{Abdikamalov:2021ues}. 

\textcolor{blue}{${\tt Afe}$}: Iron abundance of the material of the accretion disk in units of the Solar iron abundance.

\textcolor{blue}{${\tt refl\_frac}$}: Reflection fraction, namely the relative strength between the reflection component and the continuum form the corona~\cite{Dauser:2016yuj}. If it is positive, the output of the model is the reflection component and the continuum from the corona. If it is frozen to a negative number, the output of the model is only the reflection component.

\textcolor{blue}{${\tt logN}$}: Logarithm of the electron density of the disk, where the electron density is expressed in units of cm$^{-3}$. This parameter appears in the flavors that employ the table of ${\tt xillverD}$. In ${\tt relxilldgrad\_nk}$, it is the logarithm of the electron density at the inner edge of the disk~\cite{Abdikamalov:2021ues}.

\textcolor{blue}{${\tt kTe}$}: Coronal temperature in keV. This parameter appears in the flavors that employ the table of ${\tt xillverCp}$.

\textcolor{blue}{${\tt xi\_index}$}: This parameter appears in the models in which the ionization parameter profile is described by a power law and corresponds to the ionization index~\cite{Abdikamalov:2021rty}.

\textcolor{blue}{${\tt logN\_index}$}: This parameter appears in the models in which the electron density profile is described by a power law and corresponds to the electron density index~\cite{Abdikamalov:2021ues}.

\textcolor{blue}{${\tt r\_ring}$}: Radius of the ring-like corona in units of gravitational radii. This parameter appears in ${\tt relxillring\_nk}$ and ${\tt rellinering\_nk}$~\cite{Riaz:2020svt}.

\textcolor{blue}{${\tt rdisk\_in}$}: Inner radius of the disk-like corona in units of gravitational radii. This parameter appears in ${\tt relxilldisk\_nk}$ and ${\tt rellinedisk\_nk}$~\cite{Riaz:2020svt}.

\textcolor{blue}{${\tt rdisk\_out}$}: Outer radius of the disk-like corona in units of gravitational radii. This parameter appears in ${\tt relxilldisk\_nk}$ and ${\tt rellinedisk\_nk}$~\cite{Riaz:2020svt}.

\textcolor{blue}{${\tt norm}$}: Normalization of the component.

\subsubsection*{How to use the model}

${\tt relxill\_nk}$ requires a FITS file for the transfer functions (which can be generated by ${\tt raytransfer}$), a FITS file for the non-relativistic reflection spectrum (if we want to calculate a full reflection spectrum, while it is not necessary such a FITS file for the ${\tt relline\_nk}$ models; the FITS file for the non-relativistic reflection spectrum can be downloaded from Thomas Dauser's webpage\footnote{\href{http://www.sternwarte.uni-erlangen.de/~dauser/research/relxill/index.html}{http://www.sternwarte.uni-erlangen.de/~dauser/research/relxill/index.html}}), and a FITS file for the emissivity profile (if we do not want to use the twice broken power law emissivity profile and we assume a lamppost, ring-like, or disk-like corona; the FITS file can be generated by ${\tt blacklamp}$). The FITS files are also available upon request by contacting us at \href{mailto:relxill_nk@fudan.edu.cn}{relxill\_nk@fudan.edu.cn}. Please note that the FITS file for the transfer function is not the FITS file of the transfer functions of ${\tt nkbb}$ (see Subsection~\ref{ss-rt}).

To use ${\tt relxill\_nk}$ with ${\tt XSPEC}$, first, you have to open the file ${\tt relbase.h}$ and change the variable RELXILL\_TABLE\_PATH (line~48) to the path where the FITS files are stored\footnote{Alternatively, for a csh shell, you can run the command

{\tt setenv RELXILL\_TABLE\_PATH /home/user/data/relline\_tables/}

while, for a bash shell, the command 

{\tt export RELXILL\_TABLE\_PATH='/home/user/data/relline\_tables/'}}. To create the model, you can use the script ${\tt compile\_relxill.sh}$. The line commands are:

\noindent {\tt chmod u+r compile\_relxill.sh}

\noindent {\tt ./compile\_relxill.sh}

\noindent Last, it is necessary to load the model in ${\tt XSPEC}$. If you open ${\tt XSPEC}$ in the directory of the model, it is enough the following line command

\noindent {\tt lmod relxill\_nk .}

\noindent Otherwise, you need to specify the the path of the model

\noindent {\tt lmod relxill\_nk [path of the model]}

${\tt def\_par\_value}$ corresponds to the scaled value of the deformation parameter. To calculate the unscaled value of the deformation parameter appearing in the metric, there are two Python scripts: ${\tt unscale.py}$ and ${\tt unscale\_batch.py}$. The use of these two scripts is described in Subsection~\ref{ss-nkbb}

\subsection{${\tt raytransfer}$}\label{ss-rt}

${\tt raytransfer}$\footnote{\href{https://github.com/ABHModels/raytransfer}{https://github.com/ABHModels/raytransfer}} is a code to generate the FITS files of transfer functions for ${\tt nkbb}$ and ${\tt relxill\_nk}$~\cite{Zhou:2019fcg,Bambi:2016sac,Abdikamalov:2019yrr,Abdikamalov:2020oci}. The public version of ${\tt raytransfer}$ employs the Johannsen metric~\cite{Johannsen:2013szh}, but it is straightforward to modify the code to construct the FITS file for other black hole metrics. If you want to change the metric, there are the Mathematica script ${\tt CodeOptimization.nb}$ and the Maple script ${\tt CodeOptimization.mw}$ that provide the expressions of the metric coefficients, their derivative with respect to the radial coordinate, etc. ready to copy and paste in the files of ${\tt raytransfer}$. As discussed at the end of Subsection~\ref{ss-blackray}, the current version determines the ISCO radius by checking the orbital stability only along the radial direction; while this is enough for the vast majority of spacetimes, there are also cases in which the ISCO radius is determined by the stability of the orbit along the vertical direction.

In the directory with the code, there must exist the folder ``photons''. To launch the ray-tracing code, first you have to compile the code\footnote{The option {\tt -O3} is just to optimize the executive file and get a faster code.}

\noindent {\tt g++ main.cpp -O3 -o photon4trf}

\noindent and then you can launch it with the command line

\noindent {\tt ./photon4trf [spin] [mdot] [deformation]}

\noindent where you have to specify the values of the spin parameter ({\tt [spin]}), Eddington-scaled mass accretion rate ({\tt [mdot]}), and deformation parameter ({\tt [deformation]}). The code generates 22~plain data files, one for every inclination angle (see line~27 in ${\tt main.cpp}$). These data files are in the folder ``photons''.

Before generating the FITS file, you need to run ${\tt isco.cpp}$

\noindent {\tt g++ isco.cpp -O3 -o isco}

\noindent {\tt ./isco}
 
\noindent The output is the file ${\tt isco.dat}$. At this point, you can generate the FITS file with the script ${\tt Transfer.py}$

\noindent {\tt python3 Transfer.py}

\noindent The output is a FITS file of the transfer functions of the spacetime.

\subsubsection*{Notes}

It is normally enough to calculate the transfer function for 30~values of the black hole spin parameter, 30~values of the deformation parameter, and 22~values of the viewing angle. However, this depends on the exact spacetime metric that we want to implement in ${\tt nkbb}$ or ${\tt relxill\_nk}$ as well as on the parameter space that we want to scan in the data analysis. If we want to construct a new model only to analyze a specific source, it may be enough to construct a smaller FITS file (if we already know the properties of the source). For every point of the parameter space, the public version of ${\tt raytransfer}$ calculates the transfer function at 100~radii and 40~values of $g^*$. In the case of ${\tt nkbb}$, the disk emissivity profile does not drop very quickly, so it is necessary to calculate the transfer function from the ISCO to about $10^6$~gravitational radii; moreover, the values of $g^*$ are not evenly distributed, but it is necessary to have more points around $g^* = 0$ and 1, where the value of the derivative of the transfer function is high. For ${\tt relxill\_nk}$, normally we calculate the transfer function from the ISCO to about $10^3$~gravitational radii and the values of $g^*$ are evenly distributed.

\subsection{${\tt F}$-${\tt code}$}

${\tt F}$-${\tt code}$\footnote{\href{https://github.com/ABHModels/F-code}{https://github.com/ABHModels/F-code}} is a code to generate the FITS file of the dimensionless function $F(r)$ in Eq.~(\ref{eq-F(r)}) for ${\tt nkbb}$~\cite{Zhou:2019fcg}. The public version assumes the Johannsen metric~\cite{Johannsen:2013szh}, but it is straightforward to modify the code to work with another black hole spacetime. To launch the code

\noindent {\tt g++ main.cpp -O3 -o main}

\noindent {\tt ./main}

\noindent and you generate text files containing $F(r)$ vs $r$, see Eq.~(\ref{eq-F(r)}). To generate the FITS file to be used with ${\tt nkbb}$, there is the jupyter notebook ${\tt gen\_fits.ipynb}$.

\subsection{${\tt blacklamp}$}

${\tt blacklamp}$\footnote{\href{https://github.com/ABHModels/blacklamp}{https://github.com/ABHModels/blacklamp}} is a code to generate the FITS file of the emissivity profile for ${\tt relxill\_nk}$~\cite{Abdikamalov:2019yrr,Riaz:2020svt}. The model has three branches: {\tt lamppost}, {\tt ring}, and {\tt disk}. The codes in these three branches can calculate the emissivity profile for, respectively, a lamppost corona, a ring-like corona, and a disk-like corona. The current model assumes the Johannsen spacetime with deformation parameters $\alpha_{13}$, $\alpha_{22}$, $\alpha_{52}$, and $\epsilon_3$.

In the lamppost setup, the corona is a point-like source along the black hole spin axis. The emission is isotropic in the rest frame of the corona. There is only 1~parameter: the height of the corona, which is measured in gravitational radii. To calculate the emissivity profile, the command lines are

\noindent {\tt g++ main.cpp -O3 -o main}

\noindent {\tt ./main [spin] [height] [mdot]}

\noindent where you have to write the values of the black hole spin parameter ({\tt [spin]}), height of the corona ({\tt [height]}), and Eddington-scaled mass accretion rate ({\tt [mdot]}). The public version employs the Johannsen metric and the value of the deformation parameters are in ${\tt main.cpp}$, line~28. The output is a plain file.

In the ring-like setup, the corona is described by a ring above the accretion disk and there are 2~parameters: the height and the radius of the corona, both measured in gravitational radii. To calculate the emissivity profile, the command lines are

\noindent {\tt g++ mainloop\_ring.cpp -O3 -o main}

\noindent {\tt ./main [spin] [deformation]}

\noindent where you have to write the values of the black hole spin parameter ({\tt [spin]}) and of the deformation parameter $\alpha_{13}$ ({\tt [deformation]}). To change the values of the deformation parameters $\alpha_{22}$, $\alpha_{52}$, and $\epsilon_3$, open the file ${\tt mainloop\_ring.cpp}$ and go to line~187. The code generates 34~output files for different coronal heights and fixed coronal radius. To change the number or values of coronal heights or the value of the coronal radius, you have to modify ${\tt mainloop\_ring.cpp}$.

In the disk-like setup, the corona is described by a disk with a central hole above the accretion disk and there are 3~parameters: the height, the radius of the inner edge, and the radius of the outer edge of the corona, all measured in gravitational radii. To calculate the emissivity profile, the command lines are

\noindent {\tt g++ mainloop\_disk.cpp -O3 -o main}

\noindent {\tt ./main [spin] [deformation]}

\noindent where {\tt [spin]} is the value of the black hole spin parameter and {\tt [deformation]} is the value of the deformation parameter $\alpha_{13}$. To change the values of the deformation parameters $\alpha_{22}$, $\alpha_{52}$, and $\epsilon_3$, open the file ${\tt mainloop\_disk.cpp}$ and go to line~186. The code generates 25~output files for different coronal heights. To change the number or values of coronal heights or the values of the radii of the inner and outer edges of the corona, you have to modify ${\tt mainloop\_disk.cpp}$. If you want to change the emissivity profile of the corona, you have to modify ${\tt geomet.h}$.

To generate the FITS file for ${\tt relxill\_nk}$, there is the Python script ${\tt generatefits.py}$. The command line is

\noindent {\tt python3 generatefits.py}


\section{Results}\label{s-results}

${\tt nkbb}$ and ${\tt relxill\_nk}$ can be used in ${\tt XSPEC}$ to analyze X-ray spectra of black holes. The public versions on GitHub are designed to test the Kerr black hole hypothesis, namely if the spacetime around astrophysical black holes is described by the Kerr solution as expected in the framework of conventional physics. ${\tt nkbb}$ and ${\tt relxill\_nk}$ can employ a deformed Kerr spacetime, in which there is a deformation parameter quantifying possible deviations from the Kerr solution: if the deformation parameter vanishes, the metric reduces exactly to the Kerr solution of General Relativity, while a non-vanishing deformation parameter produces a deformed Kerr spacetime. One can analyze X-ray spectra of black holes with ${\tt nkbb}$ and/or ${\tt relxill\_nk}$, keeping the deformation parameter free in the fit, and try to constrain the value of the deformation parameter from observations. ${\tt relxill\_nk}$ has been extensively used to test the Kerr black hole hypothesis with both stellar-mass black holes in X-ray binaries~\cite{Tripathi:2020yts,Xu:2018lom,Liu:2019vqh,Zhang:2019ldz,Tripathi:2020dni,Zhang:2021ymo} and supermassive black holes in active galactic nuclei~\cite{Cao:2017kdq,Tripathi:2018lhx,Tripathi:2018bbu,Tripathi:2019bya,Liu:2020fpv}. ${\tt nkbb}$ has only been used to test the Kerr black hole hypothesis with stellar-mass black holes in X-ray binaries~\cite{Tripathi:2020qco,Zhang:2021ymo,Tripathi:2021rqs} because the temperature of the disk depends on the black hole mass and the spectrum turns out to peak in the soft X-ray band for stellar-mass black holes and in the UV band for supermassive black holes; in the second case, an accurate measurement of the spectrum  is impossible because of dust absorption. Most of these tests of the Kerr hypothesis have followed an agnostic strategy, employing metrics with generic deviations from the Kerr spacetime and that are not solutions of specific field equations. However, ${\tt relxill\_nk}$ has also been used to constrain specific gravity models, including Einstein-Maxwell-dilaton-axion gravity~\cite{Tripathi:2021rwb}, Kaluza-Klein gravity~\cite{Zhu:2020cfn}, asymptotically safe quantum gravity~\cite{Zhou:2020eth}, conformal gravity~\cite{Zhou:2018bxk,Zhou:2019hqk}, and regular black hole metrics~\cite{Riaz:2022rlx,Riaz:2023yde}. With minor modifications, ${\tt relxill\_nk}$ can also be used to test the Weak Equivalence Principle near black holes~\cite{Roy:2021pns} (i.e., whether the trajectory of any freely-falling test-particle is independent of its internal structure and composition).

\begin{figure}[t]
\begin{center}
\includegraphics[width=15.0cm]{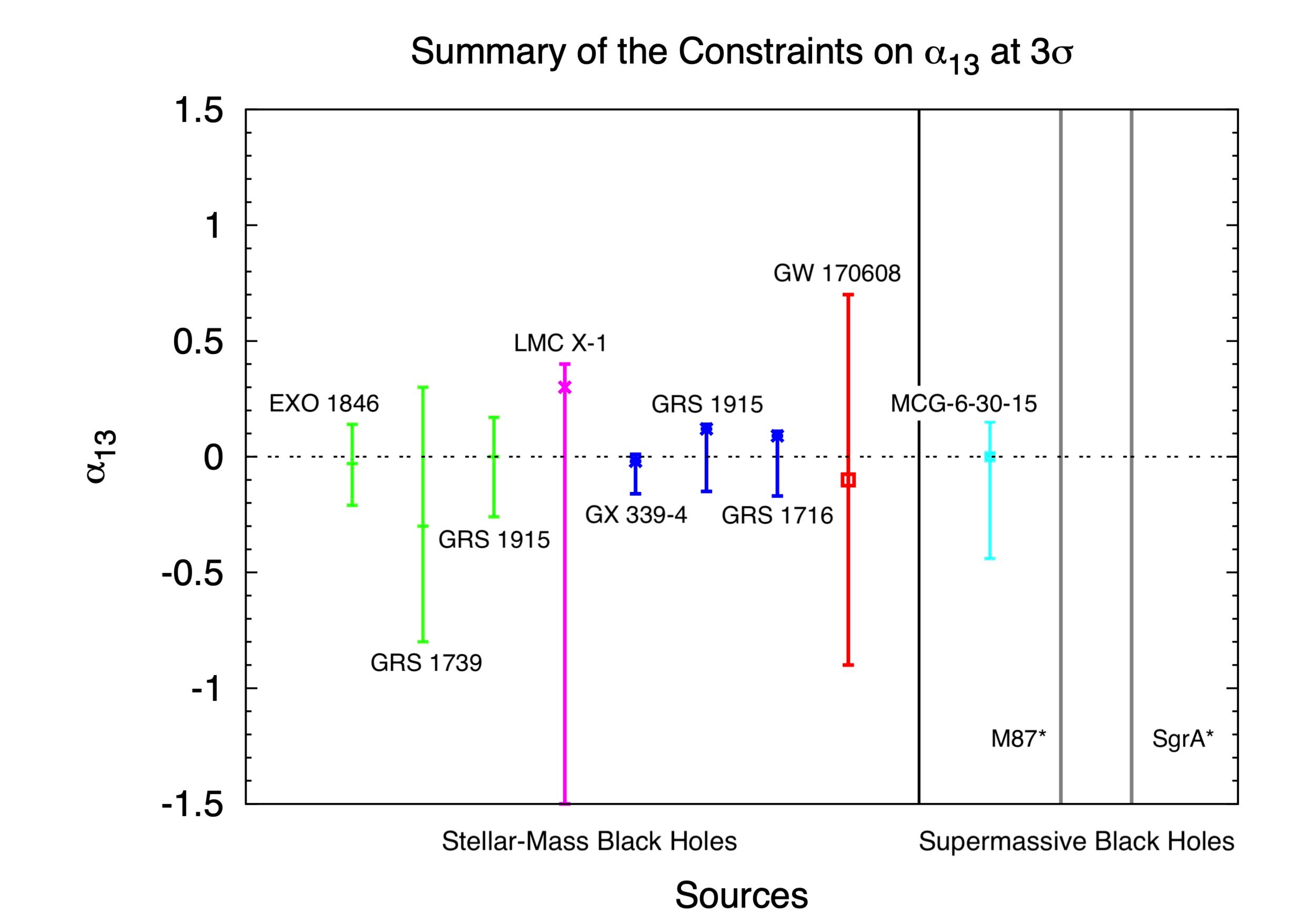}
\end{center}
\vspace{-0.5cm}
\caption{Summary of the 3-$\sigma$ constraints on the deformation parameter $\alpha_{13}$ of the Johannsen spacetime. Stellar-mass black holes: constraints from the analysis of reflection spectra with ${\tt relxill\_nk}$ (in green), thermal spectra with ${\tt nkbb}$ (in magenta), combined analysis of reflection and thermal spectra with ${\tt relxill\_nk}$ and ${\tt nkbb}$ (in blue), and gravitational waves (in red). Supermassive black holes: constraints from the analysis of reflection spectra with ${\tt relxill\_nk}$ (in cyan) and black hole imaging (in gray). See the text for more details. Figure adapted from~\cite{ref-summary} \label{f-summary}}
\end{figure}

The interested reader can find a detailed summary of the current results with ${\tt nkbb}$ and ${\tt relxill\_nk}$ in the recent reviews in Refs.~\cite{Bambi:2022dtw,Bambi:2023mca}. Here we simply discuss the current constraints on the deformation parameter $\alpha_{13}$ of the Johannsen metric from X-ray (by using ${\tt nkbb}$ and/or ${\tt relxill\_nk}$), gravitational waves, and black hole imaging. The deformation parameter $\alpha_{13}$ has no special properties, but it has been extensively used to test the Kerr hypothesis and publications in the literature have reported its constraints from different techniques. Fig.~\ref{f-summary} shows the current 3-$\sigma$ constraints on $\alpha_{13}$ ($\alpha_{13} = 0$ corresponds to the Kerr solution and is the horizontal dotted line in Fig.~\ref{f-summary}):

\noindent $\bullet$ The measurements of $\alpha_{13}$ in green correspond to the three best measurements from the analysis of the reflection features in the spectra of stellar-mass black holes with ${\tt relxill\_nk}$~\cite{Tripathi:2020yts,Zhang:2019ldz}. There are other constraints reported in the literature, but they are weaker and therefore are not reported in this plot. 

\noindent $\bullet$ The measurement in magenta corresponds to the constraint from the analysis of the thermal component of the stellar-mass black hole in LMC~X-1 with ${\tt nkbb}$~\cite{Tripathi:2020qco}. The constraint is weak because there is a strong parameter degeneracy between the spin parameter and the deformation parameter $\alpha_{13}$. 

\noindent $\bullet$ The measurements in blue correspond to three special cases in which it was possible to constrain $\alpha_{13}$ by using ${\tt nkbb}$ and ${\tt relxill\_nk}$ together~\cite{Tripathi:2020dni,Zhang:2021ymo,Tripathi:2021rqs}. The resulting measurements represent the most stringent constraints on $\alpha_{13}$ as of now, which shows that even if the analysis of the thermal component alone cannot provide a good test of the Kerr hypothesis, if combined with the analysis of the reflection features it can help to get stronger constraints. 

\noindent $\bullet$ The measurement in red corresponds to the best constraint from gravitational wave data and comes from the event GW170608~\cite{Shashank:2021giy,Cardenas-Avendano:2019zxd}. The other gravitational wave events provide weaker constraints on $\alpha_{13}$ and therefore are not shown in the plot. 

\noindent $\bullet$ The measurement in cyan corresponds to the best constraint from the analysis of the reflection features in the spectra of supermassive black holes with ${\tt relxill\_nk}$ and comes from the analysis of simultaneous \textsl{NuSTAR} and \textsl{XMM-Newton} observations of MCG--6--30--15~\cite{Tripathi:2018lhx}. In general, stellar-mass black holes can provide better constraints than supermassive black holes, because they are brighter. However, MCG--6--30--15 is quite a bright active galactic nucleus, the iron line in its reflection spectrum is very prominent and broadened, and the quality of the \textsl{NuSTAR} and \textsl{XMM-Newton} data used to measure $\alpha_{13}$ are exceptionally good: in the end, the constraint on $\alpha_{13}$ from MCG--6--30--15 is comparable to the best constraints from stellar-mass black holes.  

\noindent $\bullet$ The constraints in gray are inferred from mm VLBI data of M87$^*$ and SgrA$^*$ from the Event Horizon Telescope Collaboration~\cite{EventHorizonTelescope:2020qrl,EventHorizonTelescope:2022xqj}. These constraints are definitively weaker than those currently possible with X-ray and gravitational wave data (their 3-$\sigma$ measurement extends beyond the range of $\alpha_{13}$ of the plot). They may become comparable to the constraints from X-ray and gravitational wave data after improving the angular resolution by an order of magnitude, which could be possible by having one of the telescopes of the network in space on a satellite.

As of now, the most stringent constraints on $\alpha_{13}$ come from the combined use of ${\tt nkbb}$ and ${\tt relxill\_nk}$. While it would be incorrect to conclude that X-ray tests can provide the most stringent constraints for every deformation parameter, because different deformations from the Kerr metric are associated to different relativistic effects that have a different impact on the electromagnetic and gravitational wave spectra, it is certainly true that, in general, X-ray tests can provide very competitive constraints. In the future, gravitational wave tests are thought to improve faster. However, the two methods are complementary. In general, gravitational wave tests should probe better the dynamical regime, because they can directly test the Einstein Equations. X-ray tests can probe better the interactions between the matter and the gravity sectors. For example, the presence of an interaction between the gravity and the electromagnetic sectors making photons follow non-geodesic trajectories can easily leave a signature in the X-ray spectrum without affecting any gravitational wave signal.

Last, it is important to stress that X-ray tests require an accurate choice of the sources and the observations in order to be simultaneously precise and accurate; see the discussion in Ref.~\cite{Bambi:2022dtw} for the details. The reason is that the theoretical models behind ${\tt nkbb}$ and ${\tt relxill\_nk}$ are quite simple and suitable only for very specific systems. For example, it is extremely important to select sources with thin accretion disks. If we do not do so, we can easily get very precise but not very accurate measurements, which cannot be used to test General Relativity~\cite{Riaz:2019bkv,Riaz:2019kat,Shashank:2022xyh}. The construction of reflection models suitable to analyze X-ray data of black holes with thick disks is more challenging, mainly because there are no simple and realistic models for thick disks like the Novikov-Thorne model for thin disks, and we do not plan to implement thick disk solutions in ${\tt relxill\_nk}$ in the near future.

In the case of X-ray data, we have many observations from a number of different sources and we have to select the sources and the observations suitable for testing General Relativity: in the end, we can use only a few spectra from the most recent X-ray missions. Note the remarkable difference with the other techniques. In the case of gravitational wave tests, one can certainly get the best constraints from certain observations, but the selection of the sources and the observations is not crucial: in the end, we have always two black holes in vacuum and every observation has more or less the same systematic uncertainties. On the other hand, in the case of black hole imaging, we have only two sources, M87$^*$ and SgrA$^*$, and therefore it is compulsory to understand these systems even if they are not the ideal candidates for testing General Relativity: we do not have other choices.


\section{Concluding Remarks}\label{s-fr}

${\tt nkbb}$ and ${\tt relxill\_nk}$ are the state-of-the-art tools for testing General Relativity in the strong field regime with black hole X-ray data. If we select the correct sources and observations, we can obtain precise and accurate tests of the Kerr hypothesis and, as shown in Section~\ref{s-results}, current X-ray constraints are certainly very competitive when compared to the constraints from other techniques. Despite that, the next generation of X-ray missions (e.g., \textsl{eXTP}, \textsl{Athena}, and \textsl{HEP-X}) promise to provide unprecedented high-quality data, which will certainly require more accurate synthetic spectra than those available today. Our current efforts are thus to develop ${\tt relxill\_nk}$ in view of future, higher-quality black hole X-ray data. Fore example, we will need a more sophisticated disk model than the Novikov-Thorne one, emissivity profiles from specific coronal geometries, effects currently ignored like the returning radiation will be implemented, etc.

The development of ${\tt relxill\_nk}$ with its current architecture presents the following problem. The model relies on pre-calculated quantities that are tabulated in heavy FITS files. In some cases, the size of these FITS files is already over 1~GB, which is near the maximum size for the RAM of normal laptops. We cannot construct larger FITS files. For example, with the current version of ${\tt relxill\_nk}$, we can measure only one deformation parameter at a time. If we wanted to measure simultaneously two deformation parameters, we would need to construct a FITS file for the transfer functions of about 30~GB, which could run only on special servers. Even the inclusion of the effect of the returning radiation would require a FITS file of reflection spectra much larger than the current one, as the incident radiation would have a reflection spectrum rather than a simple power law spectrum.

To solve the problem of the size of the FITS files, we are now rewriting ${\tt relxill\_nk}$ with a neural network. After the training process, the neural network should be able to predict reflection spectra for arbitrary sets of input parameters. The training process can be time consuming, but it can be done on some large computer cluster. The neural network will be made of small files and, after the training process, it will be possible to run the model on normal laptops. With a neural network, we should be able to construct more sophisticated models than those possible today based on FITS files, as we can easily increase the number of parameters of the model.


\section*{Acknowledgments}

This work was supported by the National Natural Science Foundation of China (NSFC), Grant No.~12250610185, 11973019, and 12261131497, the Shanghai Municipal Education Commission, Grant No. 2019-01-07-00-07-E00035, and the Natural Science Foundation of Shanghai, Grant No. 22ZR1403400.


\bigskip
\bigskip
\noindent {\bf DISCUSSION}

\bigskip
\noindent {\bf DHEERAJ PASHAM:} How will your results change if we change the geometry of the corona?

\bigskip 
\noindent {\bf COSIMO BAMBI:} The geometry of the corona mainly determines the emissivity profile of the reflection spectrum. If we employ a phenomenological emissivity profile, like a twice broken power law, this should be enough to approximate well most coronal geometries~\cite{Gonzalez:2017gzu}, at least for current observations. For example, in Ref.~\cite{Tripathi:2020yts} we showed that in those cases in which a lamppost emissivity profile is preferred by the fit, one can still use a broken power law emissivity profile and get consistent estimates of the model parameters. For future observations with missions like \textsl{eXTP}, \textsl{Athena}, and \textsl{HEP-X}, which promise to provide very high-quality spectra, a twice broken power law may not be enough to approximate well most coronal geometries, but this is an issue that it is still to be studied.

\vspace{0.5cm}

\bigskip
\noindent {\bf MATTEO BACHETTI:} Would it be possible to add self-consistent polarimetric predictions to the models?

\bigskip 
\noindent {\bf COSIMO BAMBI:} Yes, it would be possible. To construct the current versions of ${\tt nkbb}$ and ${\tt relxill\_nk}$, we have to solve the photon geodesic equations backwards in time from the image plane of the observer to the emission point on the disk. If we want to add the predictions of the polarization, we have to solve simultaneously the geodesic equations and the equations of parallel transport of the polarization vector.

\vspace{0.5cm}

\bigskip
\noindent {\bf ROBERTO TAVERNA:} Is self-irradiation included in your models?

\bigskip 
\noindent {\bf COSIMO BAMBI:} The self-irradiation is not included in the current versions of ${\tt nkbb}$ and ${\tt relxill\_nk}$. Concerning thermal spectra, it is shown in Ref.~\cite{Li:2004aq} that the effect of self-irradiation can be reabsorbed in the mass accretion rate: synthetic spectra with self-irradiation look like synthetic spectra without self-irradiation and a higher mass accretion rate. So the fact that our model ${\tt nkbb}$ does not include self-irradiation should not lead to systematic uncertainties in our tests of General Relativity. Concerning reflection spectra, there are some studies showing that the self-irradiation does not have a strong impact on the estimate of most parameters~\cite{Riaz:2020zqb,Riaz:2023xng}, but those studies are not conclusive because they rely on a number of assumptions. At the moment, it would not be straightforward to include self-irradiation into the model because of the problem of the size of the FITS files (see Section~\ref{s-fr}). We plan to include the effect of self-irradiation once we will have a reflection model based on a neural network. 

\bigskip
\noindent {\bf ROBERTO TAVERNA:} And how is the ionization profile calculated?

\bigskip 
\noindent {\bf COSIMO BAMBI:} In general, we assume an ionization constant over the whole disk. However, one can also fit the data with ${\tt relxillion\_nk}$ and ${\tt relxilldgrad\_nk}$, which assume a non-trivial ionization profile. In ${\tt relxillion\_nk}$, the ionization profile is modeled with a power law and the index of the ionization profile is a parameter to be determined by the fit. In ${\tt relxilldgrad\_nk}$, the electron density is modeled with a power law and the ionization profile is calculated self-consistently from the emissivity profile and the electron density. However, as shown in Ref.~\cite{Mall:2022llu}, it seems like that most observations can be fit assuming a constant ionization parameter and that the impact of a non-trivial ionization profile on the estimate of the model parameters is weak.


\begin{thebibliography}{99}

\bibitem{Einstein:1916vd}
A.~Einstein,
\emph{The foundation of the general theory of relativity},
\href{https://doi.org/10.1002/andp.19163540702}{Annalen Phys. \textbf{49}, 769-822 (1916)}.
  
\bibitem{Will:2014kxa}
C.~M.~Will,
\emph{The Confrontation between General Relativity and Experiment},
\href{https://doi.org/10.12942/lrr-2014-4}{Living Rev. Rel. \textbf{17}, 4 (2014)}
[arXiv:1403.7377 [gr-qc]]. 

\bibitem{Ferreira:2019xrr}
P.~G.~Ferreira,
\emph{Cosmological Tests of Gravity},
\href{https://doi.org/10.1146/annurev-astro-091918-104423}{Ann. Rev. Astron. Astrophys. \textbf{57}, 335-374 (2019)}
[arXiv:1902.10503 [astro-ph.CO]].
 
\bibitem{Bambi:2017khi}
C.~Bambi,
\emph{Black Holes: A Laboratory for Testing Strong Gravity}
(Springer Singapore, 2017),
\href{https://doi.org/10.1007/978-981-10-4524-0}{DOI: 10.1007/978-981-10-4524-0},
ISBN 978-981-10-4524-0. 

\bibitem{Carter:1971zc}
B.~Carter,
\emph{Axisymmetric Black Hole Has Only Two Degrees of Freedom},
\href{https://doi.org/10.1103/PhysRevLett.26.331}{Phys. Rev. Lett. \textbf{26}, 331-333 (1971)}.

\bibitem{Robinson:1975bv}
D.~C.~Robinson,
\emph{Uniqueness of the Kerr black hole},
\href{https://doi.org/10.1103/PhysRevLett.34.905}{Phys. Rev. Lett. \textbf{34}, 905-906 (1975)}.

\bibitem{Chrusciel:2012jk}
P.~T.~Chrusciel, J.~Lopes Costa and M.~Heusler,
\emph{Stationary Black Holes: Uniqueness and Beyond},
\href{https://doi.org/10.12942/lrr-2012-7}{Living Rev. Rel. \textbf{15}, 7 (2012)}
[arXiv:1205.6112 [gr-qc]].
  
\bibitem{Kerr:1963ud}
R.~P.~Kerr,
\emph{Gravitational field of a spinning mass as an example of algebraically special metrics},
\href{https://doi.org/10.1103/PhysRevLett.11.237}{Phys. Rev. Lett. \textbf{11}, 237-238 (1963)}.

\bibitem{Bambi:2008hp}
C.~Bambi, A.~D.~Dolgov and A.~A.~Petrov,
\emph{Black holes as antimatter factories},
\href{https://doi.org/10.1088/1475-7516/2009/09/013}{JCAP \textbf{09}, 013 (2009)}
[arXiv:0806.3440 [astro-ph]].

\bibitem{Bambi:2014koa}
C.~Bambi, D.~Malafarina and N.~Tsukamoto,
\emph{Note on the effect of a massive accretion disk in the measurements of black hole spins},
\href{https://doi.org/10.1103/PhysRevD.89.127302}{Phys. Rev. D \textbf{89}, 127302 (2014)}
[arXiv:1406.2181 [gr-qc]].

\bibitem{Yunes:2009hc}
N.~Yunes and F.~Pretorius,
\emph{Dynamical Chern-Simons Modified Gravity. I. Spinning Black Holes in the Slow-Rotation Approximation},
\href{https://doi.org/10.1103/PhysRevD.79.084043}{Phys. Rev. D \textbf{79}, 084043 (2009)}
[arXiv:0902.4669 [gr-qc]].

\bibitem{Kleihaus:2011tg}
B.~Kleihaus, J.~Kunz and E.~Radu,
\emph{Rotating Black Holes in Dilatonic Einstein-Gauss-Bonnet Theory},
\href{https://doi.org/10.1103/PhysRevLett.106.151104}{Phys. Rev. Lett. \textbf{106}, 151104 (2011)}
[arXiv:1101.2868 [gr-qc]].

\bibitem{Dvali:2011aa}
G.~Dvali and C.~Gomez,
\emph{Black Hole's Quantum N-Portrait},
\href{https://doi.org/10.1002/prop.201300001}{Fortsch. Phys. \textbf{61}, 742-767 (2013)}
[arXiv:1112.3359 [hep-th]].

\bibitem{Giddings:2014ova}
S.~B.~Giddings,
\emph{Possible observational windows for quantum effects from black holes},
\href{https://doi.org/10.1103/PhysRevD.90.124033}{Phys. Rev. D \textbf{90}, 124033 (2014)}
[arXiv:1406.7001 [hep-th]].

\bibitem{Mottola:2023jxl}
E.~Mottola,
\emph{Gravitational Vacuum Condensate Stars},
in \emph{Regular Black Holes: Towards a New Paradigm of Gravitational Collapse} (Ed. C. Bambi, Springer Singapore, 2023),
\href{https://doi.org/10.1007/978-981-99-1596-5_8}{DOI: 10.1007/978-981-99-1596-5\_8}
[arXiv:2302.09690 [gr-qc]].
  
\bibitem{Herdeiro:2014goa}
C.~A.~R.~Herdeiro and E.~Radu,
\emph{Kerr black holes with scalar hair},
\href{https://doi.org/10.1103/PhysRevLett.112.221101}{Phys. Rev. Lett. \textbf{112}, 221101 (2014)}
[arXiv:1403.2757 [gr-qc]].

\bibitem{Herdeiro:2016tmi}
C.~Herdeiro, E.~Radu and H.~R\'unarsson,
\emph{Kerr black holes with Proca hair},
\href{https://doi.org/10.1088/0264-9381/33/15/154001}{Class. Quant. Grav. \textbf{33}, 154001 (2016)}
[arXiv:1603.02687 [gr-qc]].

\bibitem{Bambi:2015kza}
C.~Bambi,
\emph{Testing black hole candidates with electromagnetic radiation},
\href{https://doi.org/10.1103/RevModPhys.89.025001}{Rev. Mod. Phys. \textbf{89}, 025001 (2017)}
[arXiv:1509.03884 [gr-qc]].

\bibitem{Yagi:2016jml}
K.~Yagi and L.~C.~Stein,
\emph{Black Hole Based Tests of General Relativity},
\href{https://doi.org/10.1088/0264-9381/33/5/054001}{Class. Quant. Grav. \textbf{33}, 054001 (2016)}
[arXiv:1602.02413 [gr-qc]].

\bibitem{LIGOScientific:2016lio}
B.~P.~Abbott \textit{et al.} [LIGO Scientific and Virgo],
\emph{Tests of general relativity with GW150914},
\href{https://doi.org/10.1103/PhysRevLett.116.221101}{Phys. Rev. Lett. \textbf{116}, 221101 (2016)
[erratum: Phys. Rev. Lett. \textbf{121}, 129902 (2018)]}
[arXiv:1602.03841 [gr-qc]].

\bibitem{Yunes:2016jcc}
N.~Yunes, K.~Yagi and F.~Pretorius,
\emph{Theoretical Physics Implications of the Binary Black-Hole Mergers GW150914 and GW151226},
\href{https://doi.org/10.1103/PhysRevD.94.084002}{Phys. Rev. D \textbf{94}, 084002 (2016)}
[arXiv:1603.08955 [gr-qc]].

\bibitem{LIGOScientific:2019fpa}
B.~P.~Abbott \textit{et al.} [LIGO Scientific and Virgo],
\emph{Tests of General Relativity with the Binary Black Hole Signals from the LIGO-Virgo Catalog GWTC-1},
\href{https://doi.org/10.1103/PhysRevD.100.104036}{Phys. Rev. D \textbf{100}, 104036 (2019)}
[arXiv:1903.04467 [gr-qc]].

\bibitem{Bambi:2019tjh}
C.~Bambi, K.~Freese, S.~Vagnozzi and L.~Visinelli,
\emph{Testing the rotational nature of the supermassive object M87* from the circularity and size of its first image},
\href{https://doi.org/10.1103/PhysRevD.100.044057}{Phys. Rev. D \textbf{100}, 044057 (2019)}
[arXiv:1904.12983 [gr-qc]].

\bibitem{EventHorizonTelescope:2020qrl}
D.~Psaltis \textit{et al.} [Event Horizon Telescope],
\emph{Gravitational Test Beyond the First Post-Newtonian Order with the Shadow of the M87 Black Hole},
\href{https://doi.org/10.1103/PhysRevLett.125.141104}{Phys. Rev. Lett. \textbf{125}, 141104 (2020)}
[arXiv:2010.01055 [gr-qc]].

\bibitem{Vagnozzi:2022moj}
S.~Vagnozzi, R.~Roy, Y.~D.~Tsai, L.~Visinelli, M.~Afrin, A.~Allahyari, P.~Bambhaniya, D.~Dey, S.~G.~Ghosh and P.~S.~Joshi, \textit{et al.}
\emph{Horizon-scale tests of gravity theories and fundamental physics from the Event Horizon Telescope image of Sagittarius A$^*$},
\href{https://doi.org/10.1088/1361-6382/acd97b}{Class. Quant. Grav. (in press)}
[arXiv:2205.07787 [gr-qc]].

\bibitem{Cao:2017kdq}
Z.~Cao, S.~Nampalliwar, C.~Bambi, T.~Dauser and J.~A.~Garcia,
\emph{Testing general relativity with the reflection spectrum of the supermassive black hole in 1H0707$-$495},
\href{https://doi.org/10.1103/PhysRevLett.120.051101}{Phys. Rev. Lett. \textbf{120}, 051101 (2018)}
[arXiv:1709.00219 [gr-qc]].

\bibitem{Tripathi:2018lhx}
A.~Tripathi, S.~Nampalliwar, A.~B.~Abdikamalov, D.~Ayzenberg, C.~Bambi, T.~Dauser, J.~A.~Garcia and A.~Marinucci,
\emph{Toward Precision Tests of General Relativity with Black Hole X-Ray Reflection Spectroscopy},
\href{https://doi.org/10.3847/1538-4357/ab0e7e}{Astrophys. J. \textbf{875}, 56 (2019)}
[arXiv:1811.08148 [gr-qc]].

\bibitem{Tripathi:2020yts}
A.~Tripathi, Y.~Zhang, A.~B.~Abdikamalov, D.~Ayzenberg, C.~Bambi, J.~Jiang, H.~Liu and M.~Zhou,
\emph{Testing General Relativity with NuSTAR data of Galactic Black Holes},
\href{https://doi.org/10.3847/1538-4357/abf6cd}{Astrophys. J. \textbf{913}, 79 (2021)}
[arXiv:2012.10669 [astro-ph.HE]].

\bibitem{Bambi:2022dtw}
C.~Bambi,
\emph{Testing Gravity with Black Hole X-Ray Data},
\href{https://doi.org/10.48550/arXiv.2210.05322}{arXiv:2210.05322} [gr-qc].

\bibitem{Bambi:2023mca}
C.~Bambi,
\emph{X-Ray Tests of General Relativity with Black Holes},
\href{https://doi.org/10.3390/sym15061277}{Symmetry \textbf{15}, 1277 (2023)}
[arXiv:2305.10715 [gr-qc]].

\bibitem{Timmes:1995kp}
F.~X.~Timmes, S.~E.~Woosley and T.~A.~Weaver,
\emph{The Neutron star and black hole initial mass function},
\href{https://doi.org/10.1086/176778}{Astrophys. J. \textbf{457}, 834 (1996)}
[arXiv:astro-ph/9510136 [astro-ph]].

\bibitem{Olejak:2019pln}
A.~Olejak, K.~Belczynski, T.~Bulik and M.~Sobolewska,
\emph{Synthetic catalog of black holes in the Milky Way},
\href{https://doi.org/10.1051/0004-6361/201936557}{Astron. Astrophys. \textbf{638}, A94 (2020)}
[arXiv:1908.08775 [astro-ph.SR]].

\bibitem{Bambi:2020jpe}
C.~Bambi, L.~W.~Brenneman, T.~Dauser, J.~A.~Garcia, V.~Grinberg, A.~Ingram, J.~Jiang, H.~Liu, A.~M.~Lohfink and A.~Marinucci, \textit{et al.}
\emph{Towards Precision Measurements of Accreting Black Holes Using X-Ray Reflection Spectroscopy},
\href{https://doi.org/10.1007/s11214-021-00841-8}{Space Sci. Rev. \textbf{217}, 65 (2021)}
[arXiv:2011.04792 [astro-ph.HE]].

\bibitem{Bambi:2021chr}
C.~Bambi,
\emph{Testing General Relativity with black hole X-ray data: a progress report},
\href{https://doi.org/10.1007/s40065-021-00336-y}{Arab. J. Math. \textbf{11}, 81-90 (2022)}
[arXiv:2106.04084 [gr-qc]].

\bibitem{Ross:2005dm}
R.~R.~Ross and A.~C.~Fabian,
\emph{A Comprehensive range of x-ray ionized reflection models},
\href{https://doi.org/10.1111/j.1365-2966.2005.08797.x}{Mon. Not. Roy. Astron. Soc. \textbf{358}, 211-216 (2005)}
[arXiv:astro-ph/0501116 [astro-ph]].

\bibitem{Garcia:2010iz}
J.~Garcia and T.~Kallman,
\emph{X-ray reflected spectra from accretion disk models. I. Constant density atmospheres},
\href{https://doi.org/10.1088/0004-637X/718/2/695}{Astrophys. J. \textbf{718}, 695 (2010)}
[arXiv:1006.0485 [astro-ph.HE]].

\bibitem{Bambi:2016sac}
C.~Bambi, A.~Cardenas-Avendano, T.~Dauser, J.~A.~Garcia and S.~Nampalliwar,
\emph{Testing the Kerr black hole hypothesis using X-ray reflection spectroscopy},
\href{https://doi.org/10.3847/1538-4357/aa74c0}{Astrophys. J. \textbf{842}, 76 (2017)}
[arXiv:1607.00596 [gr-qc]].

\bibitem{Lindquist:1966igj}
R.~W.~Lindquist,
\emph{Relativistic transport theory},
\href{https://doi.org/10.1016/0003-4916(66)90207-7}{Annals Phys. \textbf{37}, 487-518 (1966)}.

\bibitem{Bambi:2013nla}
C.~Bambi,
\emph{Can the supermassive objects at the centers of galaxies be traversable wormholes? The first test of strong gravity for mm/sub-mm very long baseline interferometry facilities},
\href{https://doi.org/10.1103/PhysRevD.87.107501}{Phys. Rev. D \textbf{87}, 107501 (2013)}
[arXiv:1304.5691 [gr-qc]].

\bibitem{Novikov:1973kta}
I.~D.~Novikov and K.~S.~Thorne,
\emph{Astrophysics of black holes} 
in ``Black holes (Les astres occlus)'' (Eds. C. DeWitt and B. S. DeWitt, Gordon and Breach, New York, 1973)

\bibitem{Page:1974he}
D.~N.~Page and K.~S.~Thorne,
\emph{Disk-Accretion onto a Black Hole. Time-Averaged Structure of Accretion Disk},
\href{https://doi.org/10.1086/152990}{Astrophys. J. \textbf{191}, 499-506 (1974)}.

\bibitem{Shimura:1995nu}
T.~Shimura and F.~Takahara,
\emph{On the spectral hardening factor of the X-ray emission from accretion disks in black hole candidates},
\href{https://doi.org/10.1086/175740}{Astrophys. J. \textbf{445}, 780-788 (1995)}.

\bibitem{Nayakshin:2000vm}
S.~Nayakshin, D.~Kazanas and T.~R.~Kallman,
\emph{Thermal instability and photoionized x-ray reflection in accretion disks},
\href{https://doi.org/doi:10.1086/309054}{Astrophys. J. \textbf{537}, 833-852 (2000)}.

\bibitem{Cunningham:1975zz}
C.~T.~Cunningham,
\emph{The effects of redshifts and focusing on the spectrum of an accretion disk around a Kerr black hole},
\href{https://doi.org/10.1086/154033}{Astrophys. J. \textbf{202}, 788-802 (1975)}.

\bibitem{Abdikamalov:2020oci}
A.~B.~Abdikamalov, D.~Ayzenberg, C.~Bambi, T.~Dauser, J.~A.~Garcia, S.~Nampalliwar, A.~Tripathi and M.~Zhou,
\emph{Testing the Kerr black hole hypothesis using X-ray reflection spectroscopy and a thin disk model with finite thickness},
\href{https://doi.org/10.3847/1538-4357/aba625}{Astrophys. J. \textbf{899}, 80 (2020)}
[arXiv:2003.09663 [astro-ph.HE]].

\bibitem{Johannsen:2013szh}
T.~Johannsen,
\emph{Regular Black Hole Metric with Three Constants of Motion},
\href{https://doi.org/10.1103/PhysRevD.88.044002}{Phys. Rev. D \textbf{88}, 044002 (2013)}
[arXiv:1501.02809 [gr-qc]].

\bibitem{Abdikamalov:2013}
A.~B.~Abdikamalov, H.~Liu, C.~Bambi {\it et al.},
in preparation.

\bibitem{Bambi:2011vc}
C.~Bambi and E.~Barausse,
\emph{The Final stages of accretion onto non-Kerr compact objects},
\href{https://doi.org/10.1103/PhysRevD.84.084034}{Phys. Rev. D \textbf{84}, 084034 (2011)}
[arXiv:1108.4740 [gr-qc]].

\bibitem{Bambi:2013eb}
C.~Bambi and G.~Lukes-Gerakopoulos,
\emph{Testing the existence of regions of stable orbits at small radii around black hole candidates},
\href{https://doi.org/10.1103/PhysRevD.87.083006}{Phys. Rev. D \textbf{87}, 083006 (2013)}
[arXiv:1302.0565 [gr-qc]].

\bibitem{Zhou:2019fcg}
M.~Zhou, A.~B.~Abdikamalov, D.~Ayzenberg, C.~Bambi, H.~Liu and S.~Nampalliwar,
\emph{XSPEC model for testing the Kerr black hole hypothesis using the continuum-fitting method},
\href{https://doi.org/10.1103/PhysRevD.99.104031}{Phys. Rev. D \textbf{99}, 104031 (2019)}
[arXiv:1903.09782 [gr-qc]].

\bibitem{Zhou:2020koa}
M.~Zhou, A.~B.~Abdikamalov, D.~Ayzenberg, C.~Bambi, V.~Grinberg and A.~Tripathi,
\emph{Thermal spectra of thin accretion discs of finite thickness around Kerr black holes},
\href{https://doi.org/10.1093/mnras/staa1591}{Mon. Not. Roy. Astron. Soc. \textbf{496}, 497-503 (2020)}
[arXiv:2004.12589 [astro-ph.HE]].

\bibitem{xspec}
K.~A.~Arnaud,
ASP Conf. Ser. \textbf{101} (1996).

\bibitem{Abdikamalov:2019yrr}
A.~B.~Abdikamalov, D.~Ayzenberg, C.~Bambi, T.~Dauser, J.~A.~Garcia and S.~Nampalliwar,
\emph{Public Release of RELXILL\_NK: A Relativistic Reflection Model for Testing Einstein\textquoteright{}s Gravity},
\href{https://doi.org/10.3847/1538-4357/ab1f89}{Astrophys. J. \textbf{878}, 91 (2019)}
[arXiv:1902.09665 [gr-qc]].

\bibitem{Riaz:2020svt}
S.~Riaz, A.~B.~Abdikamalov, D.~Ayzenberg, C.~Bambi, H.~Wang and Z.~Yu,
\emph{Reflection Spectra of Accretion Disks Illuminated by Disk-like Coronae},
\href{https://doi.org/10.3847/1538-4357/ac3827}{Astrophys. J. \textbf{925}, 51 (2022)}
[arXiv:2012.07469 [astro-ph.HE]].

\bibitem{Abdikamalov:2021rty}
A.~B.~Abdikamalov, D.~Ayzenberg, C.~Bambi, H.~Liu and Y.~Zhang,
\emph{Implementation of a radial disk ionization profile in the relxill\_nk model},
\href{https://doi.org/10.1103/PhysRevD.103.103023}{Phys. Rev. D \textbf{103}, 103023 (2021)}
[arXiv:2101.10100 [astro-ph.HE]].

\bibitem{Abdikamalov:2021ues}
A.~B.~Abdikamalov, D.~Ayzenberg, C.~Bambi, H.~Liu and A.~Tripathi,
\emph{A Reflection Model with a Radial Disk Density Profile},
\href{https://doi.org/10.3847/1538-4357/ac3237}{Astrophys. J. \textbf{923}, 175 (2021)}
[arXiv:2108.00375 [astro-ph.HE]].

\bibitem{Dauser:2013xv}
T.~Dauser, J.~Garcia, J.~Wilms, M.~Bock, L.~W.~Brenneman, M.~Falanga, K.~Fukumura and C.~S.~Reynolds,
\emph{Irradiation of an Accretion Disc by a Jet: General Properties and Implications for Spin Measurements of Black Holes},
\href{https://doi.org/10.1093/mnras/sts710}{Mon. Not. Roy. Astron. Soc. \textbf{430}, 1694 (2013)}
[arXiv:1301.4922 [astro-ph.HE]].

\bibitem{Garcia:2013oma}
J.~Garcia, T.~Dauser, C.~S.~Reynolds, T.~R.~Kallman, J.~E.~McClintock, J.~Wilms and W.~Eikmann,
\emph{X-ray reflected spectra from accretion disk models. III. A complete grid of ionized reflection calculations},
\href{https://doi.org/10.1088/0004-637X/768/2/146}{Astrophys. J. \textbf{768}, 146 (2013)}
[arXiv:1303.2112 [astro-ph.HE]].

\bibitem{Garcia:2013lxa}
J.~Garc\'\i{}a, T.~Dauser, A.~Lohfink, T.~R.~Kallman, J.~Steiner, J.~E.~McClintock, L.~Brenneman, J.~Wilms, W.~Eikmann and C.~S.~Reynolds, \textit{et al.}
\emph{Improved Reflection Models of Black-Hole Accretion Disks: Treating the Angular Distribution of X-rays},
\href{https://doi.org/10.1088/0004-637X/782/2/76}{Astrophys. J. \textbf{782}, no.2, 76 (2014)}
[arXiv:1312.3231 [astro-ph.HE]].

\bibitem{Zdziarski:1996wq}
A.~A.~Zdziarski, W.~N.~Johnson and P.~Magdziarz,
\emph{Broad-band gamma-ray and x-ray spectra of ngc 4151 and their implications for physical processes and geometry},
\href{https://doi.org/10.1093/mnras/283.1.193}{Mon. Not. Roy. Astron. Soc. \textbf{283}, 193 (1996)}
[arXiv:astro-ph/9607015 [astro-ph]].

\bibitem{Zdziarski:1998}
A.~A.~Zdziarski, C.~Done and D.~A.~Smith,
\emph{The 1989 May outburst of the soft X-ray transient GS 2023+338 (V404 Cyg)},
\href{https://doi.org/10.1046/j.1365-8711.1999.02885.x}{Mon. Not. Roy. Astron. Soc. \textbf{309}, 561 (1999)}
[arXiv:astro-ph/9904304  [astro-ph]].

\bibitem{Dauser:2016yuj}
T.~Dauser, J.~Garc\'\i{}a, D.~J.~Walton, W.~Eikmann, T.~Kallman, J.~Wilms and J.~McClintock,
\emph{Normalizing a relativistic model of X-ray reflection - Definition of the reflection fraction and its implementation in relxill},
\href{https://doi.org/10.1051/0004-6361/201628135}{Astron. Astrophys. \textbf{590}, A76 (2016)}
[arXiv:1601.03771 [astro-ph.HE]].

\bibitem{Xu:2018lom}
Y.~Xu, S.~Nampalliwar, A.~B.~Abdikamalov, D.~Ayzenberg, C.~Bambi, T.~Dauser, J.~A.~Garcia and J.~Jiang,
\emph{A Study of the Strong Gravity Region of the Black Hole in GS 1354\textendash{}645},
\href{https://doi.org/10.3847/1538-4357/aadb9d}{Astrophys. J. \textbf{865}, 134 (2018)}
[arXiv:1807.10243 [gr-qc]].

\bibitem{Liu:2019vqh}
H.~Liu, A.~B.~Abdikamalov, D.~Ayzenberg, C.~Bambi, T.~Dauser, J.~A.~Garcia and S.~Nampalliwar,
\emph{Testing the Kerr hypothesis using X-ray reflection spectroscopy with NuSTAR data of Cygnus X-1 in the soft state},
\href{https://doi.org/10.1103/PhysRevD.99.123007}{Phys. Rev. D \textbf{99}, 123007 (2019)}
[arXiv:1904.08027 [gr-qc]].

\bibitem{Zhang:2019ldz}
Y.~Zhang, A.~B.~Abdikamalov, D.~Ayzenberg, C.~Bambi and S.~Nampalliwar,
\emph{Tests of the Kerr hypothesis with GRS 1915+105 using different RELXILL flavors},
\href{https://doi.org/10.3847/1538-4357/ab4271}{Astrophys. J. \textbf{884}, 147 (2019)}
[arXiv:1907.03084 [gr-qc]].

\bibitem{Tripathi:2020dni}
A.~Tripathi, A.~B.~Abdikamalov, D.~Ayzenberg, C.~Bambi, V.~Grinberg and M.~Zhou,
\emph{Testing the Kerr Black Hole Hypothesis with GX 339\textendash{}4 by a Combined Analysis of Its Thermal Spectrum and Reflection Features},
\href{https://doi.org/10.3847/1538-4357/abccbd}{Astrophys. J. \textbf{907}, 31 (2021)}
[arXiv:2010.13474 [astro-ph.HE]].

\bibitem{Zhang:2021ymo}
Z.~Zhang, H.~Liu, A.~B.~Abdikamalov, D.~Ayzenberg, C.~Bambi and M.~Zhou,
\emph{Testing the Kerr Black Hole Hypothesis with GRS 1716-249 by Combining the Continuum Fitting and the Iron-line Methods},
\href{https://doi.org/10.3847/1538-4357/ac350e}{Astrophys. J. \textbf{924}, 72 (2022)}
[arXiv:2106.03086 [astro-ph.HE]].

\bibitem{Tripathi:2018bbu}
A.~Tripathi, S.~Nampalliwar, A.~B.~Abdikamalov, D.~Ayzenberg, J.~Jiang and C.~Bambi,
\emph{Testing the Kerr nature of the supermassive black hole in Ark 564},
\href{https://doi.org/10.1103/PhysRevD.98.023018}{Phys. Rev. D \textbf{98}, 023018 (2018)}
[arXiv:1804.10380 [gr-qc]].

\bibitem{Tripathi:2019bya}
A.~Tripathi, J.~Yan, Y.~Yang, Y.~Yan, M.~Garnham, Y.~Yao, S.~Li, Z.~Ding, A.~B.~Abdikamalov and D.~Ayzenberg, \textit{et al.}
\emph{Constraints on the Spacetime Metric around Seven \textquotedblleft{}Bare\textquotedblright{} AGNs Using X-Ray Reflection Spectroscopy},
\href{https://doi.org/10.3847/1538-4357/ab0a00}{Astrophys. J. \textbf{874}, 135 (2019)}
[arXiv:1901.03064 [gr-qc]].

\bibitem{Liu:2020fpv}
H.~Liu, H.~Wang, A.~B.~Abdikamalov, D.~Ayzenberg and C.~Bambi,
\emph{Reflection features in the X-ray spectrum of Fairall 9 and implications for tests of general relativity},
\href{https://doi.org/10.3847/1538-4357/ab917a}{Astrophys. J. \textbf{896}, 160 (2020)}
[arXiv:2004.11542 [gr-qc]].

\bibitem{Tripathi:2020qco}
A.~Tripathi, M.~Zhou, A.~B.~Abdikamalov, D.~Ayzenberg, C.~Bambi, L.~Gou, V.~Grinberg, H.~Liu and J.~F.~Steiner,
\emph{Testing general relativity with the stellar-mass black hole in LMC X-1 using the continuum-fitting method},
\href{https://doi.org/10.3847/1538-4357/ab9600}{Astrophys. J. \textbf{897}, 84 (2020)}
[arXiv:2001.08391 [gr-qc]].

\bibitem{Tripathi:2021rqs}
A.~Tripathi, A.~B.~Abdikamalov, D.~Ayzenberg, C.~Bambi, V.~Grinberg, H.~Liu and M.~Zhou,
\emph{Testing the Kerr black hole hypothesis with the continuum-fitting and the iron line methods: the case of GRS~1915+105},
\href{https://doi.org/10.1088/1475-7516/2022/01/019}{JCAP \textbf{01}, 019 (2022)}
[arXiv:2106.10982 [astro-ph.HE]].

\bibitem{Tripathi:2021rwb}
A.~Tripathi, B.~Zhou, A.~B.~Abdikamalov, D.~Ayzenberg and C.~Bambi,
\emph{Constraints on Einstein-Maxwell dilaton-axion gravity from X-ray reflection spectroscopy},
\href{https://doi.org/10.1088/1475-7516/2021/07/002}{JCAP \textbf{07}, 002 (2021)}
[arXiv:2103.07593 [astro-ph.HE]].

\bibitem{Zhu:2020cfn}
J.~Zhu, A.~B.~Abdikamalov, D.~Ayzenberg, M.~Azreg-Ainou, C.~Bambi, M.~Jamil, S.~Nampalliwar, A.~Tripathi and M.~Zhou,
\emph{X-ray reflection spectroscopy with Kaluza-Klein black holes},
\href{https://doi.org/10.1140/epjc/s10052-020-8198-x}{Eur. Phys. J. C \textbf{80}, 622 (2020)}
[arXiv:2005.00184 [gr-qc]].

\bibitem{Zhou:2020eth}
B.~Zhou, A.~B.~Abdikamalov, D.~Ayzenberg, C.~Bambi, S.~Nampalliwar and A.~Tripathi,
\emph{Shining X-rays on asymptotically safe quantum gravity},
\href{https://doi.org/10.1088/1475-7516/2021/01/047}{JCAP \textbf{01}, 047 (2021)}
[arXiv:2005.12958 [astro-ph.HE]].

\bibitem{Zhou:2018bxk}
M.~Zhou, Z.~Cao, A.~Abdikamalov, D.~Ayzenberg, C.~Bambi, L.~Modesto and S.~Nampalliwar,
\emph{Testing conformal gravity with the supermassive black hole in 1H0707-495},
\href{https://doi.org/10.1103/PhysRevD.98.024007}{Phys. Rev. D \textbf{98}, 024007 (2018)}
[arXiv:1803.07849 [gr-qc]].

\bibitem{Zhou:2019hqk}
M.~Zhou, A.~B.~Abdikamalov, D.~Ayzenberg, C.~Bambi, L.~Modesto, S.~Nampalliwar and Y.~Xu,
\emph{Singularity-free black holes in conformal gravity: New observational constraints},
\href{https://doi.org/10.1209/0295-5075/125/30002}{EPL \textbf{125}, 30002 (2019)}
[arXiv:2003.03738 [gr-qc]].

\bibitem{Riaz:2022rlx}
S.~Riaz, S.~Shashank, R.~Roy, A.~B.~Abdikamalov, D.~Ayzenberg, C.~Bambi, Z.~Zhang and M.~Zhou,
\emph{Testing regular black holes with X-ray and GW data},
\href{https://doi.org/10.1088/1475-7516/2022/10/040}{JCAP \textbf{10}, 040 (2022)}
[arXiv:2206.03729 [gr-qc]].

\bibitem{Riaz:2023yde}
S.~Riaz, A.~B.~Abdikamalov and C.~Bambi,
\emph{Testing Regular Black Holes with X-ray data of GX~339--4},
\href{https://doi.org/10.48550/arXiv.2306.09673}{arXiv:2306.09673} [astro-ph.HE].

\bibitem{Roy:2021pns}
R.~Roy, A.~B.~Abdikamalov, D.~Ayzenberg, C.~Bambi, S.~Riaz and A.~Tripathi,
\emph{Testing the weak-equivalence principle near black holes},
\href{https://doi.org/10.1103/PhysRevD.104.044001}{Phys. Rev. D \textbf{104}, 044001 (2021)}
[arXiv:2103.08978 [astro-ph.HE]].

\bibitem{ref-summary}
\href{https://cosimobambi.github.io/research.html}{https://cosimobambi.github.io/research.html}

\bibitem{Shashank:2021giy}
S.~Shashank and C.~Bambi,
\emph{Constraining the Konoplya-Rezzolla-Zhidenko deformation parameters III: Limits from stellar-mass black holes using gravitational-wave observations},
\href{https://doi.org/10.1103/PhysRevD.105.104004}{Phys. Rev. D \textbf{105}, 104004 (2022)}
[arXiv:2112.05388 [gr-qc]].

\bibitem{Cardenas-Avendano:2019zxd}
A.~Cardenas-Avendano, S.~Nampalliwar and N.~Yunes,
\emph{Gravitational-wave versus X-ray tests of strong-field gravity},
\href{https://doi.org/10.1088/1361-6382/ab8f64}{Class. Quant. Grav. \textbf{37}, 135008 (2020)}
[arXiv:1912.08062 [gr-qc]].
 
\bibitem{EventHorizonTelescope:2022xqj}
K.~Akiyama \textit{et al.} [Event Horizon Telescope],
\emph{First Sagittarius A* Event Horizon Telescope Results. VI. Testing the Black Hole Metric},
\href{https://doi.org/10.3847/2041-8213/ac6756}{Astrophys. J. Lett. \textbf{930}, L17 (2022)}.

\bibitem{Riaz:2019bkv}
S.~Riaz, D.~Ayzenberg, C.~Bambi and S.~Nampalliwar,
\emph{Reflection spectra of thick accretion discs},
\href{https://doi.org/10.1093/mnras/stz3022}{Mon. Not. Roy. Astron. Soc. \textbf{491}, 417-426 (2020)}
[arXiv:1908.04969 [astro-ph.HE]].

\bibitem{Riaz:2019kat}
S.~Riaz, D.~Ayzenberg, C.~Bambi and S.~Nampalliwar,
\emph{Modeling bias in supermassive black hole spin measurements},
\href{https://doi.org/10.3847/1538-4357/ab89ab}{Astrophys. J. \textbf{895}, 61 (2020)}
[arXiv:1911.06605 [astro-ph.HE]].

\bibitem{Shashank:2022xyh}
S.~Shashank, S.~Riaz, A.~B.~Abdikamalov and C.~Bambi,
\emph{Testing Relativistic Reflection Models with GRMHD Simulations of Accreting Black Holes},
\href{https://doi.org/10.3847/1538-4357/ac9128}{Astrophys. J. \textbf{938}, 53 (2022)}
[arXiv:2207.11526 [astro-ph.HE]].

\bibitem{Gonzalez:2017gzu}
A.~G.~Gonzalez, D.~R.~Wilkins and L.~C.~Gallo,
\emph{Probing the geometry and motion of AGN coronae through accretion disc emissivity profiles},
\href{https://doi.org/10.1093/mnras/stx2080}{Mon. Not. Roy. Astron. Soc. \textbf{472}, 1932-1945 (2017)}
[arXiv:1708.03205 [astro-ph.HE]].

\bibitem{Li:2004aq}
L.~X.~Li, E.~R.~Zimmerman, R.~Narayan and J.~E.~McClintock,
\emph{Multi-temperature blackbody spectrum of a thin accretion disk around a Kerr black hole: Model computations and comparison with observations},
\href{https://doi.org/10.1086/428089}{Astrophys. J. Suppl. \textbf{157}, 335-370 (2005)}
[arXiv:astro-ph/0411583 [astro-ph]].

\bibitem{Riaz:2020zqb}
S.~Riaz, M.~L.~Szanecki, A.~Nied\'zwiecki, D.~Ayzenberg and C.~Bambi,
\emph{Impact of the returning radiation on the analysis of the reflection spectra of black holes},
\href{https://doi.org/10.3847/1538-4357/abe2a3}{Astrophys. J. \textbf{910}, 49 (2021)}
[arXiv:2006.15838 [astro-ph.HE]].

\bibitem{Riaz:2023xng}
S.~Riaz, A.~B.~Abdikamalov and C.~Bambi,
\emph{Impact of the returning radiation in current tests of the Kerr black hole hypothesis using X-ray reflection spectroscopy},
\href{https://doi.org/10.48550/arXiv.2303.12581}{arXiv:2303.12581} [astro-ph.HE].
  
\bibitem{Mall:2022llu}
G.~Mall, A.~Tripathi, A.~B.~Abdikamalov and C.~Bambi,
\emph{Impact of ionization and electron density gradients in X-ray reflection spectroscopy measurements},
\href{https://doi.org/10.1093/mnras/stac3102}{Mon. Not. Roy. Astron. Soc. \textbf{517}, 5721-5733 (2022)}
[arXiv:2206.03478 [astro-ph.HE]].

  
\end{thebibliography}
\end{document}